# Linear coupling between lightwaves in metamaterials enables lossless artificial magnetism

Kosmas L. Tsakmakidis[1], Marek S. Wartak[1,2], Durga P. Aryal[1], Ortwin Hess[1]

[1]*Advanced Technology Institute and Department of Physics, Faculty of Engineering and Physical Sciences, University of Surrey, Guildford, GU2 7XH, United Kingdom.*
[2]*Department of Physics and Computer Science, Wilfrid Laurier University, Waterloo, Ontario N2L 3C5, Canada.*

**A well-known principle in optical physics states that power can never be exchanged between two light waves propagating inside a homogeneous medium when the medium response is strictly linear[1]. Power exchange between light waves usually occurs with the aid of nonlinearity[2-5]. A typical example is nonlinear optical parametric amplification where net energy can flow from a high-frequency (high-energy) light wave (pump) to a lower-frequency signal wave, leading to amplification of the signal wave[1]. Here, we show that this limitation of ordinary media can be overcome using suitably engineered materials, known as metamaterials, which in the recent past have enabled a variety of extraordinary applications, unattainable using conventional materials, such as 'perfect' lenses[6], 'invisibility' cloaks[7] and 'trapped rainbows' (ref 8). We introduce a blueprint for magnetic metamaterials that enables, in a *totally linear* fashion, coherent and *constructive* active power flow from one light wave to another (second) light wave. Our analysis reveals that the magnetic field component of the second wave can now be *amplified* inside the magnetic metamaterial. Importantly, we show that the amplification of the magnetic field occurs in frequency regions where the real part of the metamaterial's effective permeability is *negative* (*metamaterial* magnetism). This type of magnetic amplification, which we call '*metamaterial parametric amplification*' (MPA) is**

**similar in consequence to nonlinear two-beam coupling or optical parametric amplification but is linear and uniquely occurs in metamaterials. The scheme is extremely resilient to the presence of dissipative losses and can be implemented across the entire electromagnetic spectrum (from radio up to optical frequencies) using existing, mature metamaterial technology. It allows harnessing *perfectly lossless or active*, *linear metamaterial* magnetism over a *continuous range of frequencies* in a specified direction *without* having to use extraneous gain (e.g., stimulated emission) or nonlinear media to compensate for dissipative losses.**

Macroscopic composite materials exhibiting unusual electromagnetic properties, known as metamaterials (MMs) have enabled a multitude of applications[6-9] beyond the reach of ordinary electromagnetic media. However, at present, the performance of metamaterials is limited by the occurrence of dissipative losses (light absorption), which occasionally can be up to tens of dB/wavelength (ref 10). Clearly, if these exceptional materials are to find their way towards practical applications, this issue has to be adequately addressed.

To this end, a number of diverse strategies have been proposed[11]. A first approach relies on the use of gain media to compensate for the losses that originate from the presence of metallic elements in the 'meta-molecules' (ref 12). This approach is very promising and it has, in fact, been theoretically shown that it can lead to zero-loss metamaterials[13], even over a broad but finite bandwidth[14]. However, it may not always be possible to find a suitable gain medium to provide the necessary gain at the desired frequency regime. A second interesting scheme makes use of negatively reflecting/refractive interfaces to reproduce the features of bulk negative-index metamaterials[15]. Upon being negatively refracted at the engineered interfaces, light is allowed to propagate through lossless dielectric materials, such as air, thereby avoiding high-attenuation regions. This strategy relies on the use of nonlinear media and therefore requires intense incident light. More recently, a series of works have deployed concepts such as electromagnetically induced transparency (EIT) to overcome metamaterial losses[16-19]. However, as recognised in these works[17,18], only *reflection* losses are therein compensated (the internal, random *scattering* losses are zero for a perfectly periodic, defect-free MM crystal with deep-subwavelength unit cells), i.e. one essentially achieves a better impedance matching to the surrounding medium. Thus, the analogy to a 'true' EIT scheme wherein *absorption/dissipation* losses are overcome is not a strict one and, accordingly, when light passes (owing to the improved impedance matching) through these EIT-MM structures the actual absorption/dissipation optical losses increase[17,18] – or a low-dissipative-loss transparency window occurs in regions where the real part of the effective electromagnetic parameters is *positive*[19].

In this work we introduce a design paradigm that allows for constructing *intrinsically* lossless or *active* resonant-type magnetic metamaterials (with $\mathrm{Re}\{\mu\} < 0$), i.e. *linear* metamaterials in which the mechanism for eliminating *dissipative* losses and inducing an active response occurs inherently at the unit-cell level *without* requiring extraneous gain and/or nonlinear media. Let us start by remarking that, in general, it is possible to overcome the dissipative losses occurring in a medium without resorting to optical gain media (stimulated emission) by, e.g., deploying naturally occurring nonlinear media that can facilitate two-beam coupling (TBC) 'gain'[2-4] or optical parametric amplification (OPA)[1,20,21] – but not stimulated emission. For instance, in OPA two light beams are coupled through the $\chi^{(2)}$-nonlinear response of the medium, resulting in a net energy flow from the high-frequency (high-energy) light beam (pump) to the lower-frequency signal beam. In fact, OPA has lately been shown to enable compensation of *dissipative* losses even in MMs[21]. Unfortunately, however, such a scheme requires large field intensities, while part of the energy supplied by the pump is always converted into an idler wave. Moreover, as recognised in ref 22, it might not always be feasible to match the frequency regimes of the MM and nonlinear responses of the medium as well as obtain the required phase matching between the three interacting waves.

Such a flow of power from a pump beam to a (physically separated) signal beam (leading to the amplification of the signal beam) has previously been widely assumed to occur only inside nonlinear media[1-5], but metamaterials can overcome this limitation. Indeed, in the following we are showing that *homogeneous* MMs can, in a totally *linear* fashion, enable on the atomic (i.e. 'metaparticle') level net *active power flow* between two physically separated beams, which leads to *dissipative*-loss compensation in a manner *reminiscent* of (but clearly different in many important respects to) TBC or OPA – and without any of the aforementioned practical difficulties associated with conventional OPA – *while maintaining negative* effective-$\mu$ responses for the MM medium. This type of magnetic amplification, which we call 'metamaterial parametric amplification' (MPA; not to be confused with optical parametric amplification that can occur in nonlinear media, including metamaterials[20,21]) can *only* occur (*sui generis*) in suitably designed metamaterials (see Supplementary Information, SI, section-3).

To see how MPA can occur, consider first the single-mesh, *RLC* metaparticle shown in Fig.1(a), $R$ being the equivalent lumped ohmic resistance, $L$ the equivalent lumped inductance and $C$ the equivalent lumped capacitance of a *non-bianisotropic*[23] split-ring resonator (SRR) metaparticle. A method for extracting the effective lumped-element parameters of individual SRR particles is detailed in ref 24. Let us assume that each mesh is circular with area $S = \pi r^2$, has a unit vector $\hat{\mathbf{a}} = \mathbf{z_0}$ normal to its surface ($\mathbf{z_0}$ being the unit vector along the $z$-axis) and is located at the centre of a unit cell of volume $G = a^2 \times \ell$, periodically repeated in three dimensions (see SI, section-2). The whole structure is excited by a plane wave with magnetic field component $\mathbf{H} = H_0\exp(i\omega t - i\mathbf{k}\bullet\mathbf{r})\mathbf{z_0}$, where $\mathbf{k}$ is the wavevector, $\mathbf{r}$ is the vector along the direction of the wave propagation, $\omega$ is the angular frequency and $t$ is the time. From Faradays' law, the electromotive force (emf voltage) induced in the closed mesh by the **H**-field will be $U_{emf} = -i\omega\mu_0 SH_0$, with $S$ being the surface of the $RLC$ mesh. If $I$ is the induced current circulating the $RLC$ mesh, application of Kirchhoff's second law in this circuit leads to: $U_1 = \text{abs}(U_{emf}) = i\omega\mu_0 SH_0 = I(R + i\omega L + 1/(i\omega C)]$. Assuming low-density, weakly interacting ('isolated') unit cells, the last expression, together with the relations: $\mathbf{I} = -I\boldsymbol{\varphi}_0$ and $\mathbf{M} = -(IS/G)\mathbf{z}_0$, with $\boldsymbol{\varphi}_0$ being the polar angle unit vector, leads to the following equation for the effective relative magnetic permeability of this structure: $\mu = 1 + F\ /\ [1/(\omega^2 LC) - 1 + iR/\omega L)]$, $F = S/a^2$ being the unit cell filling factor. A plot of the real and imaginary parts of $\mu$ is also shown in Fig. 1(a), from where one may see that $\text{Im}\{\mu\} < 0$, i.e. the medium is (magnetically) passive. Note that, in the present case, the imaginary part of $\mu$ corresponds solely to *dissipative* (*not reflection/scattering*) losses[25], the rate of which is determined by the $RLC$ resonator's $Q$-factor, $Q = \sqrt{L/C}/(2R)$; indeed, for $R = 0$, it is $\text{Im}\{\mu\} = 0$. It is to be noted that in this passive configuration the direction of the current $I$ circulating the electric mesh ($\mathbf{I} = -I\boldsymbol{\varphi}_0$) is such that it induces a magnetic moment **M** *opposing* the incident **H**-field, in accordance with Lentz's law (see Fig. 1(a)).

If we could devise a physically realizable scheme that could (by some means) enable reversal of the direction of the current **I** (from $-\boldsymbol{\varphi}_0$ to $+\boldsymbol{\varphi}_0$) for the *same* incident magnetic field $\mathbf{H} = +H\mathbf{z}_0$, then the direction of the induced magnetisation **M** would also be reversed (from $-\mathbf{z}_0$, to $+\mathbf{z}_0$), i.e. it would be *co-directed* to the incident **H**-field and, as such, it would be *reinforcing* the incident **H**-field. Mathematically, since the direction of **M** would be reversed, the imaginary part of the structure's effective permeability $\mu$ would also change sign, i.e. we would have a magnetically *active* structure characterised by $\text{Im}\{\mu\} > 0$.

There is a basic (but useful for our later discussions) way to accomplish this feat. Indeed, consider the configuration shown in Fig. 1(b), where we have assumed that an *additional* ac source, $U_2^{/} = +iU_2$ ($U_2 > 0$), is electrically connected to the previous *RLC* circuit within *each* unit cell. Let us assume that the source $U_2^{/}$ is oscillating with the same frequency as $U_{\mathrm{emf}}$, but has larger amplitude and *opposite* polarity (see Fig. 1(b)). Since $|U_2^{/}| > |U_{\mathrm{emf}}|$, the generated current $I$ will now be circulating along the $+\boldsymbol{\varphi}_0$ direction and, according to our previous discussion, we should expect the induced magnetic moment **M** to have a *positive* contribution to the imaginary part of the structure's effective permeability. Indeed, Kirchhoff's law of voltages for this configuration reads (see SI, section-2): $iU_2 - i\omega\mu_0 SH_0 = I(R + i\omega L + 1/(i\omega C)]$, which results in an effective magnetic permeability with: $\mathrm{Im}\{\mu\} = FR(U_2 - \omega\mu_0 SH_0)/[\ell H_0(R^2 + W^2)] > 0$, with $W = \omega L - 1/(\omega C)$. Thus, this configuration is, as expected, magnetically *active*, characterized by an effective magnetic permeability $\mu$ with $\mathrm{Im}\{\mu\} > 0$.

It is important to realise that in the configuration of Fig. 1(b) the lossless/active effective-$\mu$ behaviour of the structure occurs *not* because there are no losses but, simply, because the *additional* energy that the external source $U_2^{/}$ inserts into the system reverses the direction of the current $I$ and 'forces' the induced magnetisation **M** to *strengthen* the incident magnetic field. Indeed, it is straightforward to show that for the circuit of Fig. 1(b) it is: $\mathrm{Re}\{U_2^{/}I^*\} = |I|^2R + \mathrm{Re}\{U_1 I^*\}$ (see SI, section-2). From this equation one can infer that when the active power emitted by the external source ($\mathrm{Re}\{U_2^{/}I^*\}$) increases, so does the circulating current $I$. Accordingly, the induced magnetisation **M** also increases. Since **M** ↑↑ **H**, this process ultimately leads to *amplification* of the incident magnetic field. At the same time, however, more ohmic losses (heat) are generated in the structure, owing to the increase in the magnitude of the circulating current $I$. Thus, not only are (dissipative) losses present in this structure, but they also increase the more the magnetic field is amplified inside it. For a given time-averaged active power provided by the external source, the simultaneous amplification of the magnetic field and the generation of heat inside the structure do not, of course, violate energy conservation, simply because the magnetic field is (equal to) the current $I$ per unit length. Accordingly, if with the present scheme the magnetic field is to be amplified, the current *I has* to increase – a process that also, unavoidably, increases the heat generated inside the structure. The generated heat can be diminished by suitable cooling without, of course, affecting the afore-described amplification of the magnetic field inside this structure.

By all means the configurations of Fig. 1(b) can allow for the construction of an *active* MM exhibiting *negative* relative permeability $\mu$. However, the realisation of such a system would, clearly, be challenging since it would require an external ac source to be connected to the metaparticle (*RLC* circuit) in *each* unit cell. A neat way to overcome this difficulty, is to realize that the circuit of Fig. 1(b) can effectively be modelled by considering a *second RLC* mesh (with elements $R_2$, $L_2$, $C_2$) electrically connected to the first mesh (with elements $R_1$, $L_1$, $C_1$). The elements of the second mesh can, e.g., be determined following a reverse '*Thévenin equivalent circuit*' (TEC) process, or they can be arbitrary, i.e. new $R_2$, $L_2$, $C_2$ elements unrelated to those obtained with a reverse TEC process. In each case, because of the exchange of active power between the two meshes (see below), the effect that mesh-2 has on mesh-1 can be modelled by considering an ac source (with some complex total impedance) connected as a 'load' to mesh-1, thereby replicating the active MM configuration of Fig. 1(b).

To see this equivalence, let us imagine that the previous magnetic field $\mathbf{H} = H_0\exp(i\omega t - i\mathbf{k}\bullet\mathbf{r})\hat{\mathbf{e}}_0$ is now incident to a *pair* of the two electrically connected meshes, such that $\hat{\mathbf{e}}_\mathbf{0}$ is parallel to $\hat{\mathbf{a}}$, where $\hat{\mathbf{a}}$ is the unit vector normal to the plane of the two meshes. From Faradays' law, the electromotive sources (voltages) induced in the two closed meshes by the **H**-field will be $V_p = V_{emf,p} = -i\omega\mu_0 S_p H_0$, $p = 1, 2$, with $S_p$ being the surface of each mesh. From the well-known 'mesh current method', one may determine the currents circulating in each loop (see Fig. 1(c)), as $I_p = \det(G_p)/\det(G)$, where:

$$G = \begin{bmatrix} R_1 + i\omega L_1 - i/\omega C_1 - i/\omega C & i/\omega C \\ i/\omega C & R_2 + i\omega L_2 - i/\omega C_2 - i/\omega C \end{bmatrix}, \qquad (1a)$$

$$G_1 = \begin{bmatrix} V_1 & i/\omega C \\ V_2 & R_2 + i\omega L_2 - i/\omega C_2 - i/\omega C \end{bmatrix}, \quad G_2 = \begin{bmatrix} R_1 + i\omega L_1 - i/\omega C_1 - i/\omega C & V_1 \\ -i/\omega C & V_2 \end{bmatrix} \qquad (1b)$$

with $R_p$, $L_p$, $C_p$ ($p$ = 1, 2) being respectively the resistance, inductance and capacitance of the $p^{th}$ mesh, $C$ the capacitance of their common branch, and 'det' designating the matrix determinant. Having calculated the currents $I_p$ circulating in each mesh, we can determine the total (emanating from *both* meshes) magnetic dipole moment per unit volume (magnetisation), $M$, as $M = (S_1I_1 + S_2I_2)/G$, where we again assume that the meshes at the centres of neighbouring cubes are sufficiently apart from each other, so that they can be regarded as weakly interacting ('isolated') (see also SI, section-6).

For this new MM structure and for the case where the areas of the two meshes are equal ($S_2 = S_1$), Figs. 2(a)-(e) report the real/imaginary parts of the structure's effective $\mu$, the corresponding figure of merit (FOM = $-$ Re$\{\mu\}$/Im$\{\mu\}$), the active powers $P_1$, $P_2$ [$P_m = \mathrm{Re}\{V_m I_m^*\}$ ($m$ = 1, 2)] in both meshes, as well as the total active power for the system of the two meshes $P = P_1 + P_2$. The exchange of active power between the two meshes is readily inferred from Figs. 2(c)-(e). There are, indeed, regions wherein $P_1$ *or* $P_2$ become negative, but *never* simultaneously (which would violate the conservation of energy). In the region where $P_1 < 0$ the current in the first mesh is *reversed* (compared to the direction assumed in Fig. 1(c)), because in this frequency region the voltage across the branch '*ab*' in Fig. 1(c) exceeds $V_{\mathrm{emf},1}$ ($|V_{ab}| > |V_{\mathrm{emf},1}|$), with $V_{ab}$ being opposite polarized compared to $V_{\mathrm{emf},1}$. The corresponding is, also, true for the frequency region wherein $P_2 < 0$. These situations are, thus, precisely analogous to the active MM configuration that we examined in Fig. 1(b). Indeed, *if the magnetic moments ($m_1$, $m_2$) of the two meshes could be 'isolated' they would result in effective magnetic permeabilities exhibiting positive imaginary parts (magnetic gain) in the regions where $P_1 < 0$ or $P_2 < 0$.*

Unfortunately, however, in the case of Fig. 1(c) the two meshes lie on the same plane, and therefore their magnetic moments *have* to be added in order to obtain the structure's total magnetisation $M = m_1 + m_2$. This *always* results in a *negative* imaginary part for the structure's effective $\mu$, because the *total* active power $P = P_1 + P_2$ provided by the two emf sources $V_{\text{emf},1}$ and $V_{\text{emf},2}$ is, as shown in Fig. 2(e), *always* positive – as opposed to the individual active powers $P_1$ and $P_2$, which do exhibit regions wherein they are negative, as shown in Figs. 2(c) and (d). Thus, with this structure we can *never* obtain an active magnetic MM structure exhibiting $\text{Im}\{\mu\} > 0$. Even so, we can still obtain substantially reduced *dissipative* losses since, as shown in Fig. 2(e), there is a region (highlighted with a dashed red line) where the *total* active power $P$ reduces abruptly towards zero, signifying reduced *dissipative* losses. Indeed, as expected, in that region the imaginary part of the structure's effective $\mu$, shown in Fig. 2(a), also reduces abruptly towards zero (highlighted by a black dashed line in Fig. 2(a)), and as a result the FOM at precisely this region increases dramatically to around 987; for instance, at $f = 13.196$ GHz, we obtain $\mu \approx -4.4575 - i0.0045165$, i.e. we achieve a substantially *negative* $\text{Re}\{\mu\}$ with simultaneously excellent FOM.

The final step for obtaining an intrinsically lossless/active magnetic MM is to realise that a way to separate/isolate the magnetic moments ($m_1$, $m_2$) of the two meshes – and therefore harness effective permeabilities having *positive* imaginary parts in the regions where $P_1 < 0$ or $P_2 < 0$ – is to place the two meshes on *different* (orthogonal) planes. To see how this can result in an active MM configuration (in a specified direction), consider the case where the first mesh is placed on the *xz* plane and the second mesh on the *yz* plane, as illustrated in the inset of Fig. 3(a). Let us, further, assume that two uniform harmonic magnetic fields of equal amplitude, $H_x = H_2 = H_0\exp\{i(\mathbf{k_2}\bullet\mathbf{r_2}-\omega t)\}\,\hat{\boldsymbol{x}}_0$ and $H_y = H_1 = H_0\exp\{i(\mathbf{k_1}\bullet\mathbf{r_1}-\omega t)\}\,\hat{\boldsymbol{y}}_0$, are simultaneously incident on the structure, generating the electromotive voltages $V_1$ and $V_2$ shown in Fig. 3(a). In this particular arrangement, the magnetic field $H_1$ generates electric currents circulating in *both* meshes. As a result, the magnetic field $H_y = H_1$ generates magnetic moments on both $xz$ and $yz$ planes, i.e. it is responsible for the generation of $\mu_{yy}$ and $\mu_{xy}$. The corresponding is also true for the magnetic field $H_x = H_2$. The 'total' $\mu_x = \mu_2$ that the $x$-component of the magnetic flux density, $B_x$, experiences – $B_x = \mu_{xx}H_x + \mu_{xy}H_y = (\mu_{xx} + \mu_{xy})H_0 = \mu_x H_0$ – can, now, be calculated by finding the 'total' current $I_2$ shown in Fig. 3(a) and following the methodology for the calculation of the effective $\mu$ that was outlined in the previous examples. Figures 3(b) and (c) report the results of such a calculation of $\mu_x = \mu_2$ and $\mu_y = \mu_1$ for the case where the coupling capacitance, $C$, of the two meshes is $C = 0.1$ pF. Note that, as expected, the imaginary parts of $\mu_1$ and $\mu_2$ follow closely the variation with frequency of the active powers $P_1$ and $P_2$, respectively, which were previously studied in Figs. 2(c)-(e). Accordingly, there are now regions where either $\mu_1$ ($\mu_y$) or $\mu_2$ ($\mu_x$) become zero or even *positive*. It should, also, be noted that, owing to the passivity of the structure, $\mu_1$ and $\mu_2$ are *never* simultaneously positive (see inset in Fig. 3(b)); in fact, as expected, the *sum* $\mathrm{Im}\{\mu_1\} + \mathrm{Im}\{\mu_2\}$ is *exactly equal* to the imaginary part of $\mu$ that was shown in Fig. 2(a), i.e. for every frequency it is: $\mathrm{Im}\{\mu_1\} + \mathrm{Im}\{\mu_2\} < 0$.

In summary, we have presented a design paradigm that allows for ultralow or (uniaxially) lossless/active magnetic metamaterials. Our analysis has revealed that metamaterials, which were until now regarded very vulnerable to dissipative losses, remarkably, turn out to be capable of not only overcoming dissipative losses, but also offering *one of the most clear-cut means of generating linear magnetic gain*. The presented two-degrees-of-freedom based (2-DEG; see SI, section-7) 'metamaterial parametric amplification' is, in principle, scalable and realizable at any frequency regime, *including the optical regime* (see SI, section-4), using existing, mature metamaterial technology and does not require extraneous nonlinear or gain (e.g., stimulated emission) media to compensate for the dissipative losses. Metamaterial parametric amplification, which here was particularly discussed as a means of overcoming metamaterial dissipative losses, is a fundamentally new means of generating electromagnetic amplification with conceivable implications far beyond the realm of metamaterials. As such, it could potentially be exploited in a wealth of photonic applications requiring light amplification in regions where ordinary gain media do not exist or in quantum information science where *linear* coupling of photons is desired.

**Supplementary Information** is linked to the online version of the paper at www.nature.com/nature.

**Acknowledgements** We wish to acknowledge discussions with Martin W. McCall (Imperial College, UK) and Jeremy Allam (University of Surrey, UK). This work was supported by the UK Engineering and Physical Sciences Research Council (EPSRC). K. L. Tsakmakidis acknowledges support by the Royal Academy of Engineering and the EPSRC through a research fellowship. M. S. Wartak would like to acknowledge financial support provided by the Natural Science and Engineering Research Council of Canada (NSERC).

**Author Contributions** K.L.T. and O.H. conceived the presented idea. K.L.T. developed the theory, performed the computations and wrote a draft of the paper. M. S. W. contributed to the computations and development of the theory. D. P. A. contributed to the discussions. O.H. guided and supervised the work at all stages.

**Author Information** Reprints and permissions information is available at www.nature.com/reprints. The authors declare no competing financial interests. Correspondence and request for materials should be addressed to O.H. (O.Hess@surrey.ac.uk).

FIGURE LEGENDS

**Figure 1 | Linear equivalent circuit models describing the effective-$\mu$ behaviour of various magnetic metamaterials.** In all cases it is assumed that: the electric meshes reside at the centre of a parallelepiped of volume $G = a^2 \times \ell$ (unit cell), periodically repeated in three dimensions; the incident magnetic field $\mathbf{H} = +H\mathbf{z_0}$ and induced magnetisation **M** increase with time; and the polarity (designated with '+'/'–' pairs) of the ac sources is an instantaneous one, designating the direction along which each source remits electric current $I$. Further, in all cases, it is assumed that the magnetic metaparticles are *non-bianisotropic*[23] and that the plates of the capacitors are not perpendicular (i.e., *they do not couple*) to the incident electric field (see also SI, section-5). **a**, Passive configuration. Here, the current $I$ exits the emf source from it positive ('+') pole. The direction of the current is such that it induces a magnetic moment *opposing* the incident $H$-field. Accordingly, this structure is always characterised by $\text{Im}\{\mu\} < 0$ (passive magnetism). The inset at the top left side illustrates a typical variation with frequency of the real (blue) and imaginary (red) part of this structure's effective $\mu$. **b**, Active configuration. Here, it is assumed that, within each unit cell of the metamaterial, an external ac source $U_2^{/} = iU_2$ is electrically connected to the previous $RLC$ meta-particle. The $U_2^{/}$ source oscillates with the same frequency as the emf source, but has opposite polarity and larger amplitude compared to the emf source ($|U_2^{/}| = U_2 > |U_{\text{emf}}| = \omega\mu_0 S H_0$). It follows that the current $I$, now, circulates along the $+\boldsymbol{\varphi}_0$ direction, generating a magnetization $\mathbf{M} \uparrow\uparrow \mathbf{H}$. Thereby, this structure will be magnetically active, characterised by $\text{Im}\{\mu\} > 0$. The inset at the top left side illustrates a typical variation with frequency of the real (blue) and imaginary (red) part of this structure's effective $\mu$. The panel at the bottom left side (adapted and reprinted with permission from ref 26. Copyright 1962, American Association of Physics Teachers) illustrates an example of how the equivalent circuit could be realised using a source-loaded split-ring resonator. Here, the direction of the circulating current $I$ has been visualized by sprinkling grass seeds on the paper on which the SRR particle resides. **c**, Equivalent circuit model of two electrically

connected meshes. Here, in a specific frequency region, the effect that the second (first) mesh has on the first (second) mesh is equivalent to an ac source of opposite polarity to $V_{\mathrm{emf},1}$ ($V_{\mathrm{emf},2}$) connected, together with some complex impedance, as a load to the first (second) mesh, thereby replicating the active configuration of **b**.

**Figure 2 | Effective permeability and active powers associated with a magnetic metamaterial described by the equivalent circuit model of Fig. 1(c).** A perpendicularly incident magnetic field $H_0$ = 1 A/m induces two electromotive voltage sources, $V_1$ and $V_2$ (see Fig. 1(c)). The values of the lumped elements are: $R_1$ = 50 Ω, $R_2$ = 0.1 Ω, $C_1 = C_2$ = 10 fF, $C$ = 0.1 pF, $L_1 = L_2$ = 16 nH. The surfaces of the two meshes are $S_1 = S_2 = 4\times\pi$ mm$^2$ and both are located at the centres of unit cells of volume 25 mm$^3$. Variation with frequency of the: **a**, Real and imaginary part of effective $\mu$. **b**, Corresponding figure of merit. **c**, Active power $P_1$. **d**, Active power $P_2$. **e**, Total active power $P = P_1 + P_2$.

**Figure 3 | Lossless metamaterial magnetism and linear exchange of optical power via 'metamaterial parametric amplification'.** **a**, To arrive at a design that can allow for harnessing intrinsically lossless metamaterial magnetism, we use the configuration of Fig. 1(c), but now we place – in each unit cell – one mesh on the *xz* plane and the other mesh on the *yz* plane (see top left inset). This topology allows for isolating and harnessing the *active* metamaterial responses of the individual meshes that occur in the frequency regions where $P_1 < 0$ (Fig. 2(c)) or $P_2 < 0$ (Fig. 2(d)). We use two separate, incident light beams of the same amplitude and oscillation frequency: One (shown in green) whose magnetic field oscillates perpendicularly to the second mesh, i.e. parallel to the *x*-axis, and a second beam (shown in red) whose magnetic field oscillates perpendicularly to the first mesh, i.e. parallel to the *y*-axis. All the other electromagnetic and geometric parameters are those of Fig. 2. Variation with

frequency of the: **b**, Real and imaginary part of the effective relative permeability along the *x*-axis ($\mu_x$). The inset illustrates the variation with frequency of the imaginary part of $\mu_x$ (red) and $\mu_y$ (green), from where one can observe that in the region where $\mu_y$ ($\mu_x$) > 0 it is, also, $\mu_x$ ($\mu_y$) < 0. **c**, Real and imaginary part of the effective relative permeability along the *y* direction ($\mu_y$). The inset illustrates the variation with frequency of the real part of $\mu_x$ (green) and $\mu_y$ (blue).

**FIGURES**

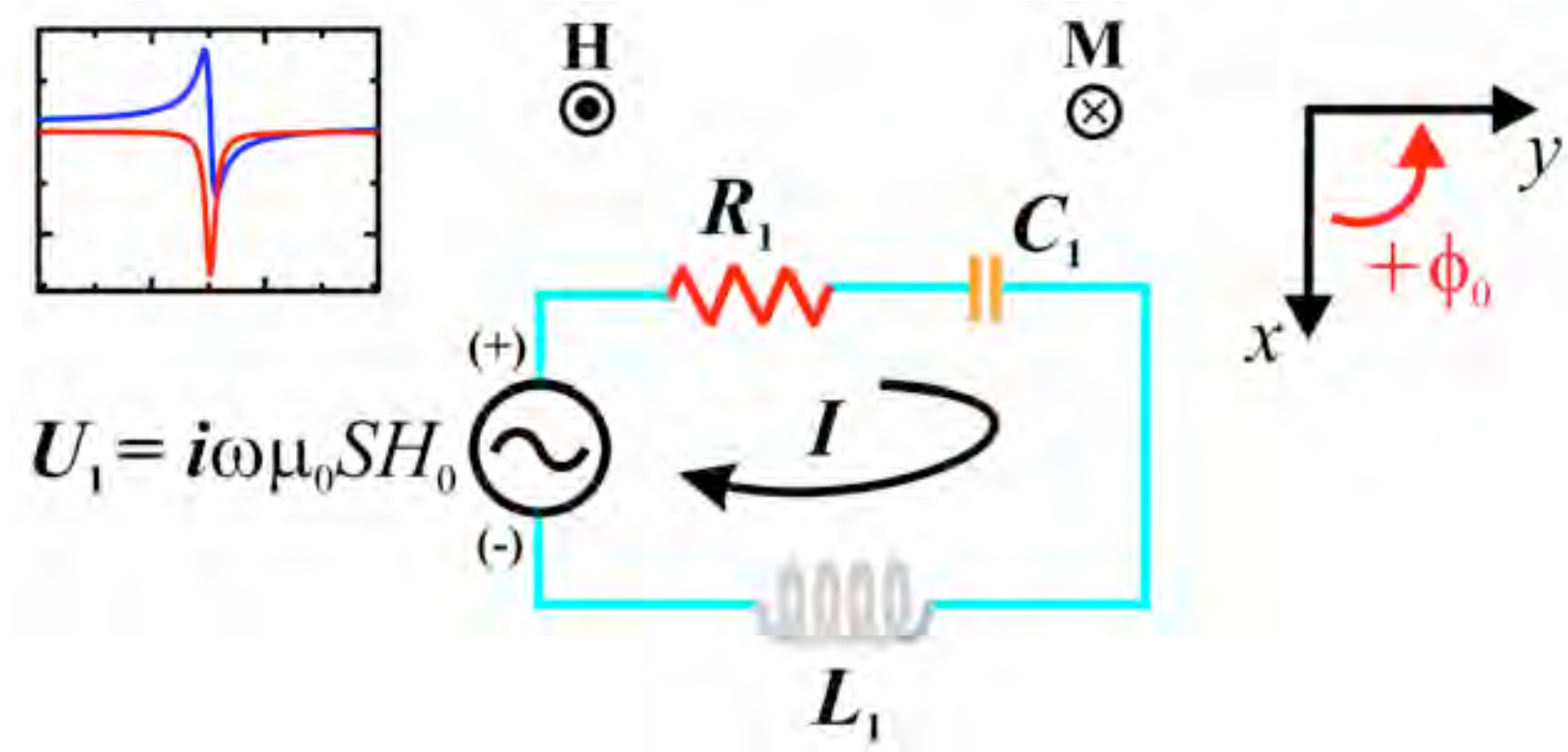

**Figure 1(a)**

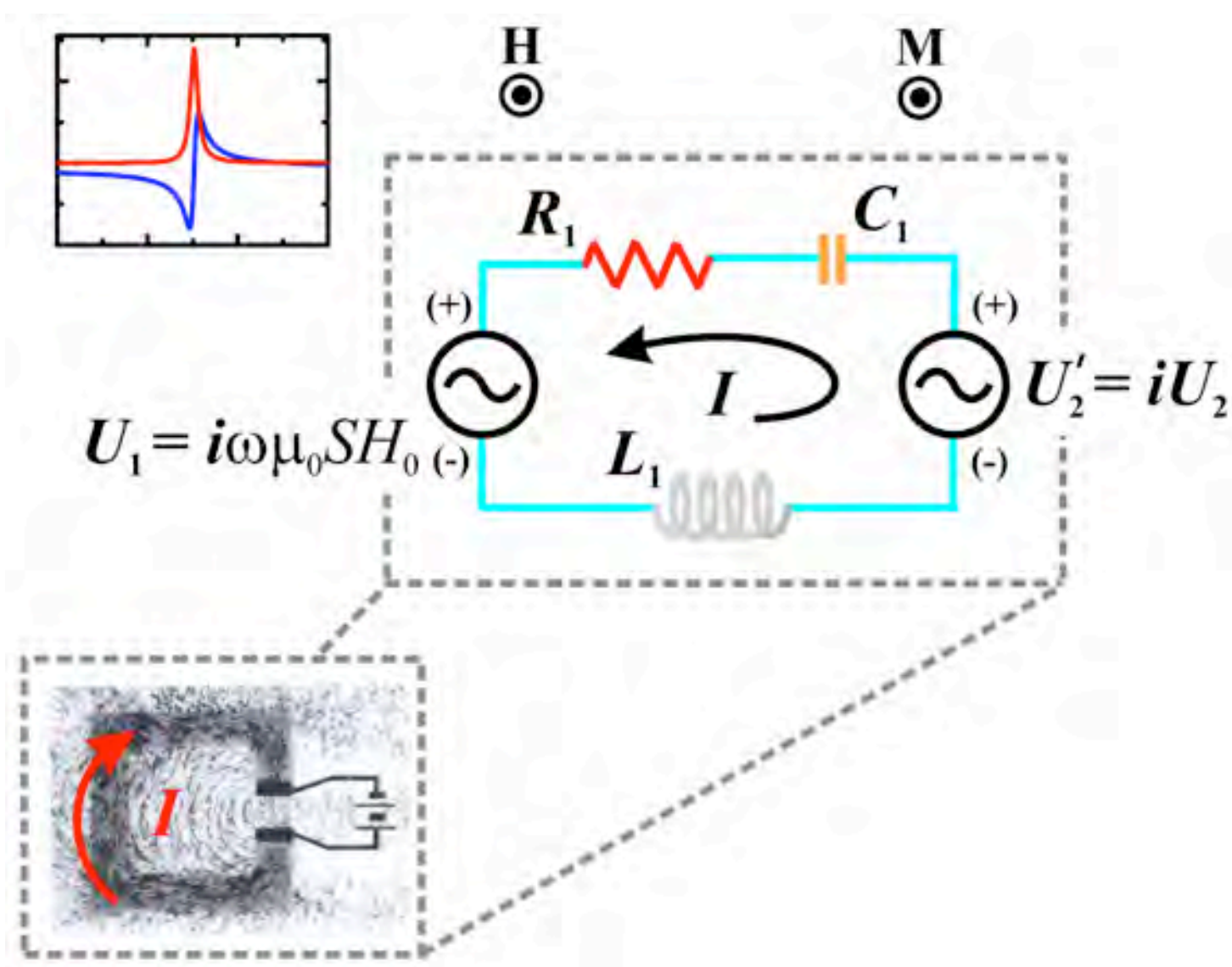

**Figure 1(b)**

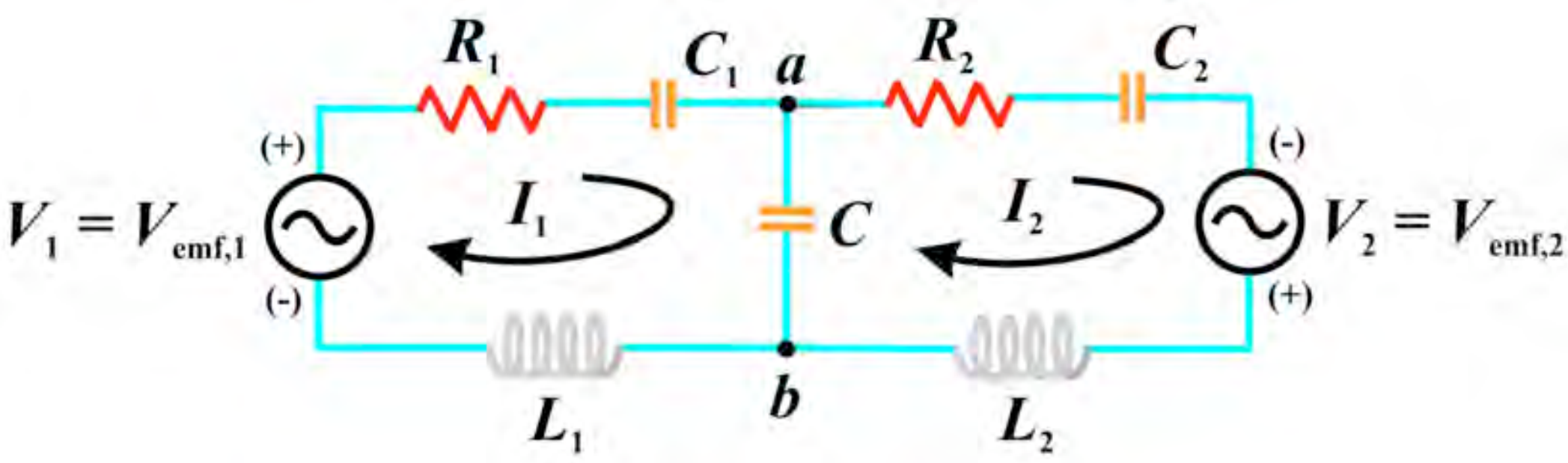

**Figure 1(c)**

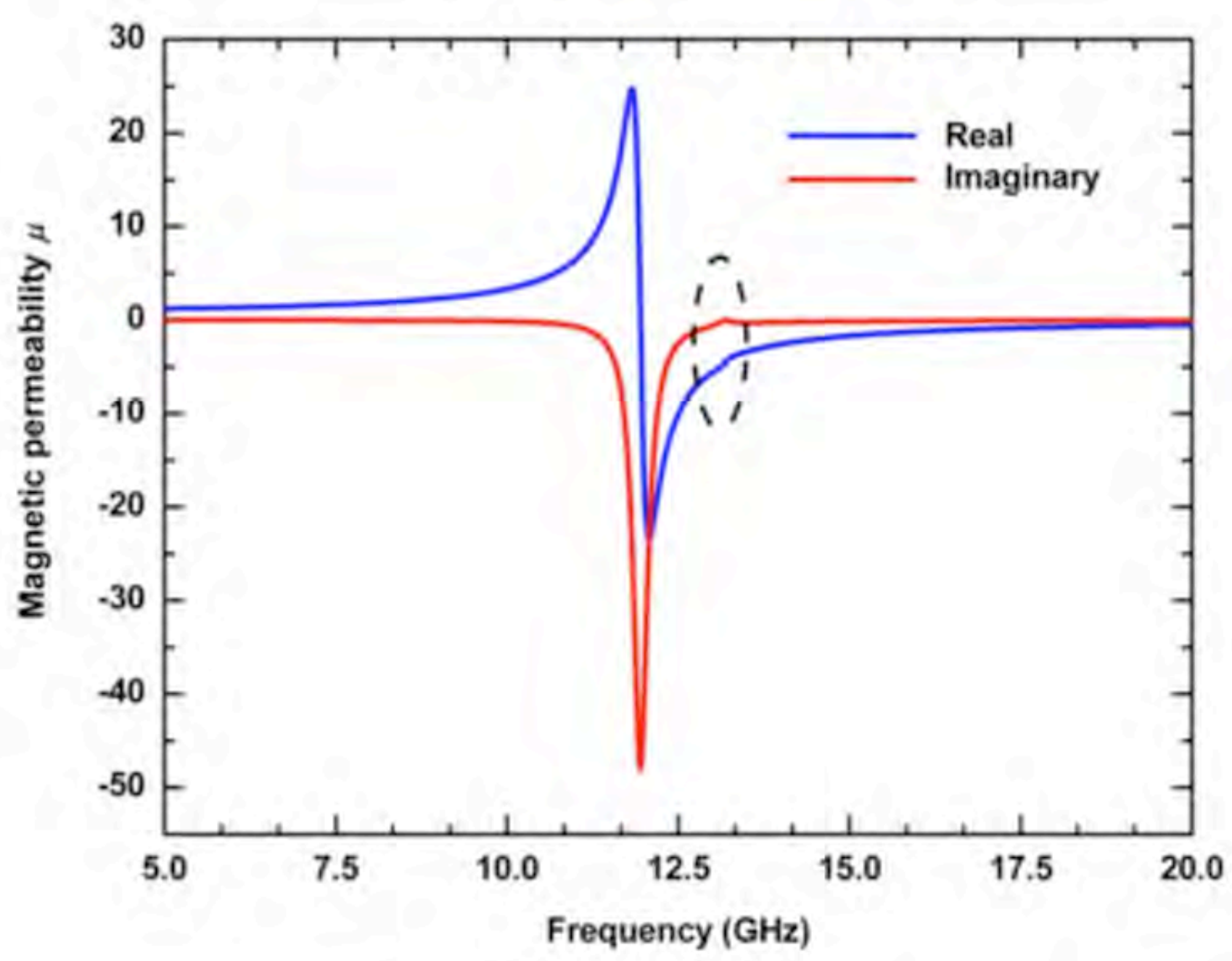

**Figure 2(a)**

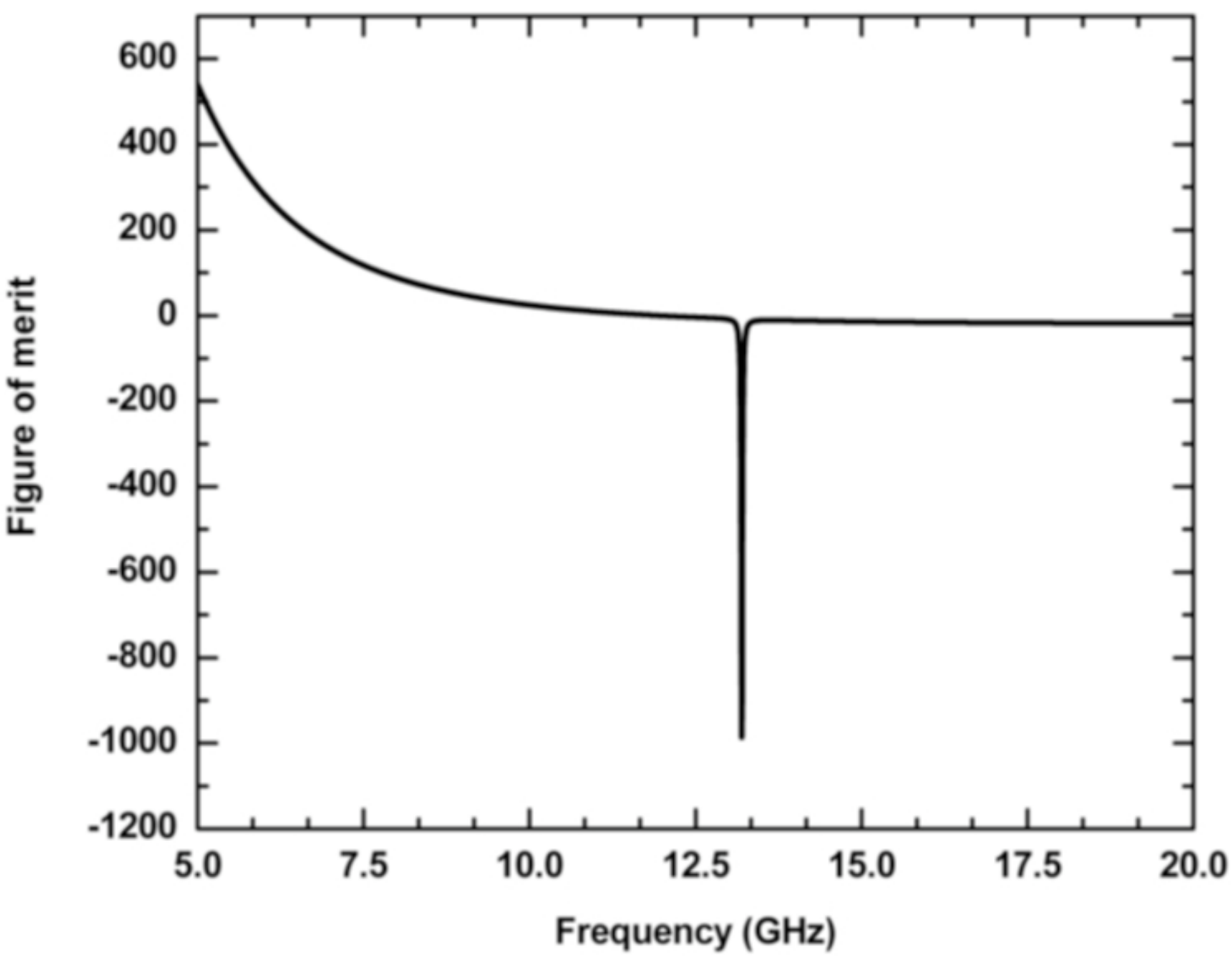

**Figure 2(b)**

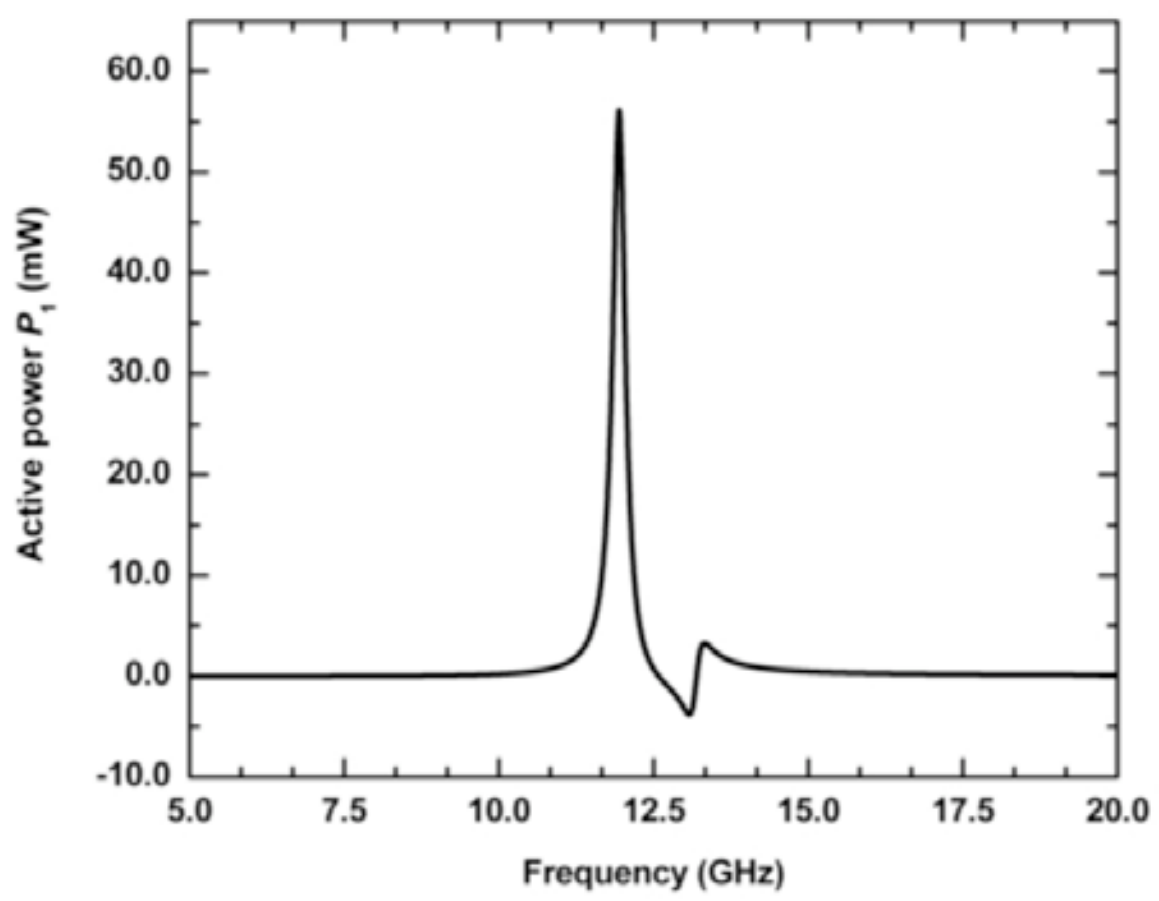

**Figure 2(c)**

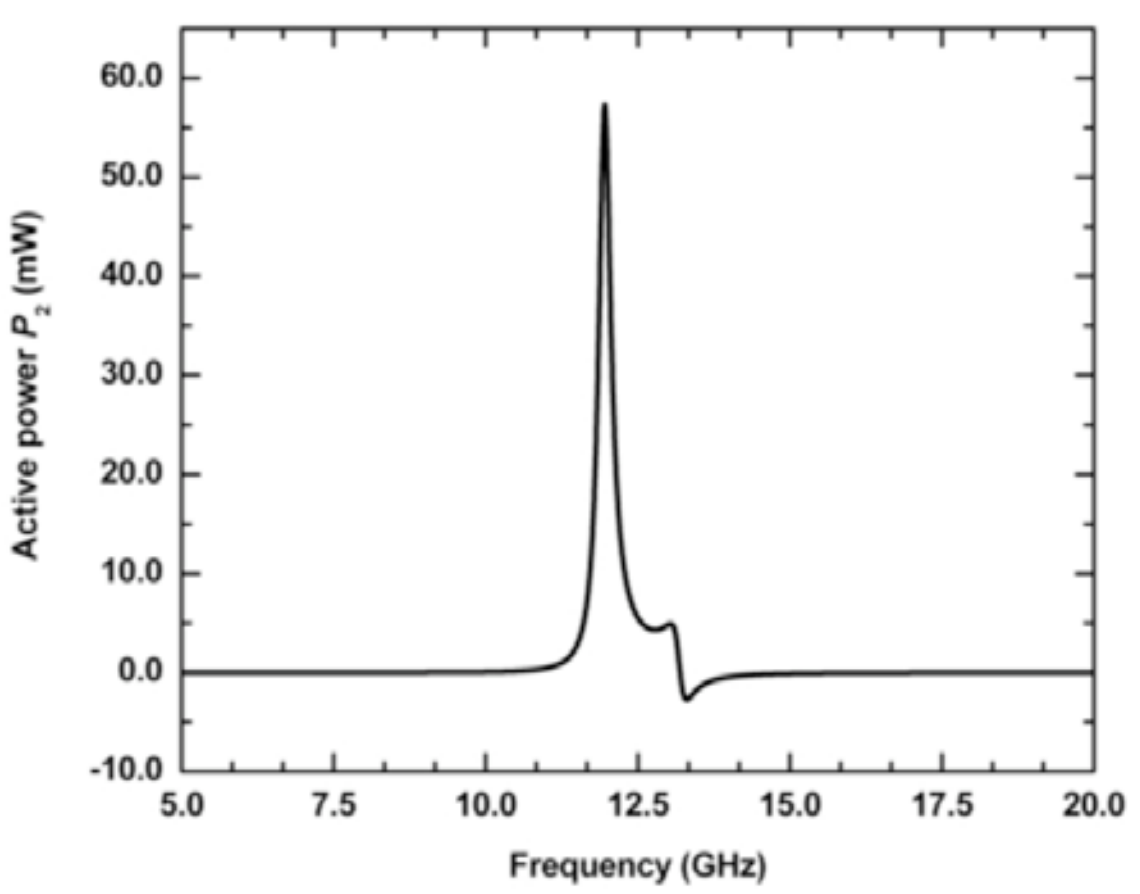

**Figure 2(d)**

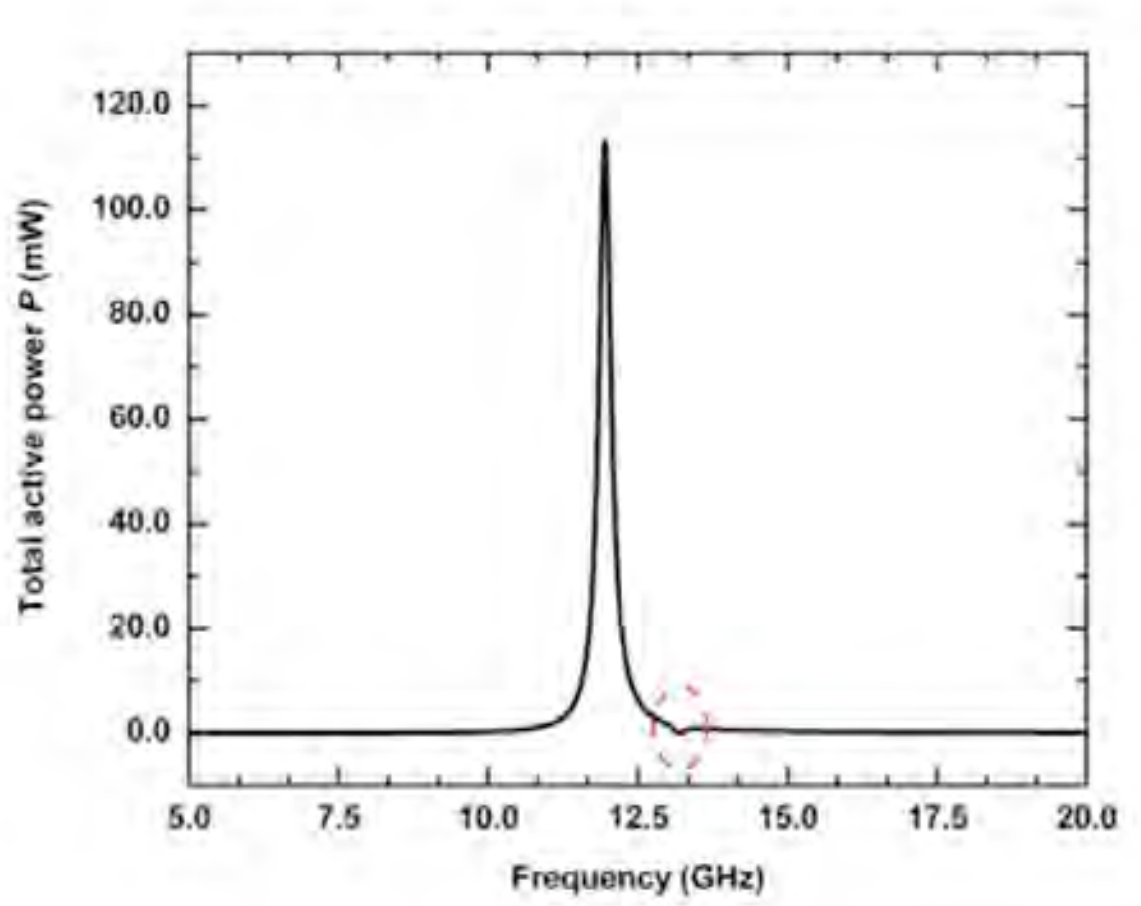

**Figure 2(e)**

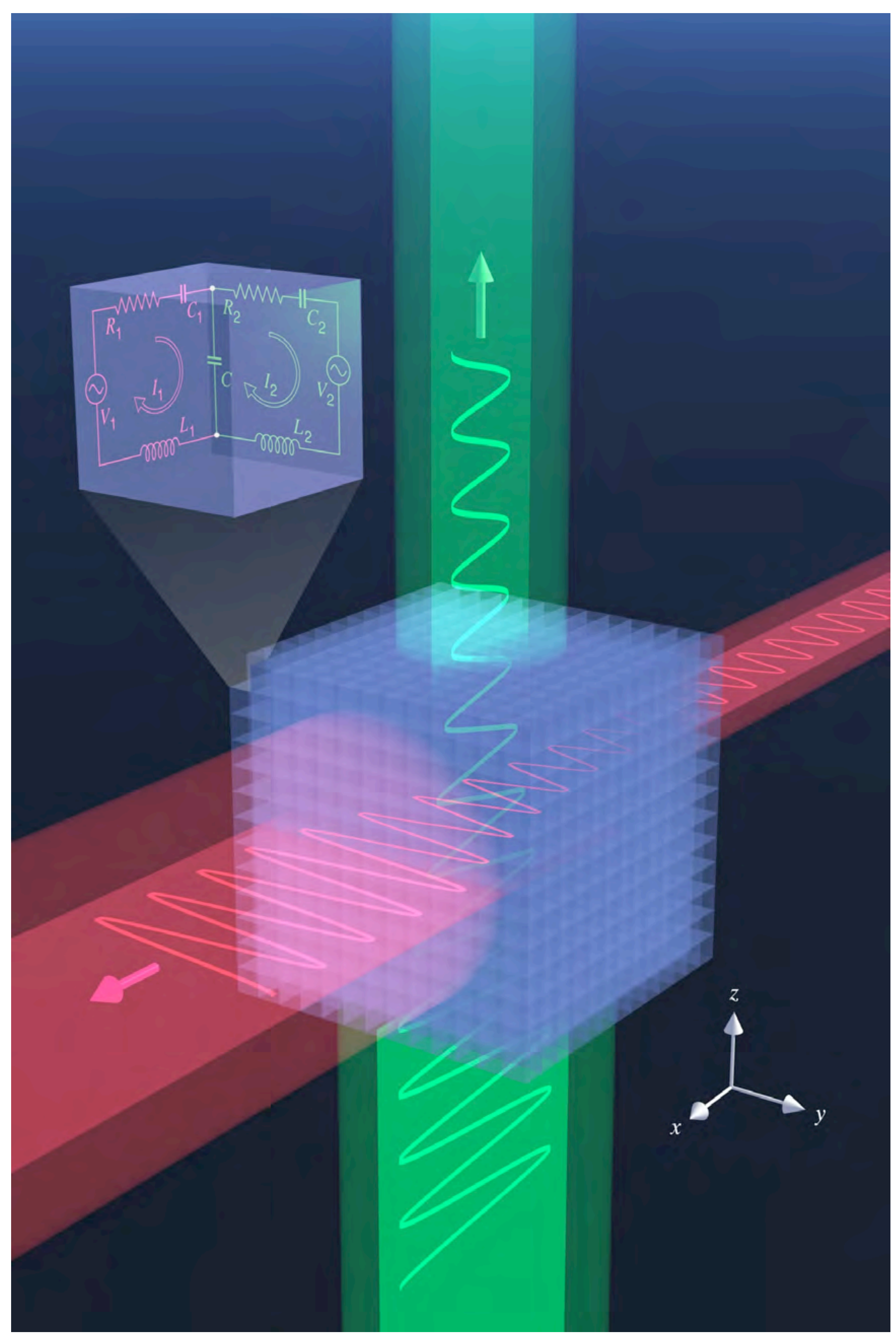

**Figure 3(a)**

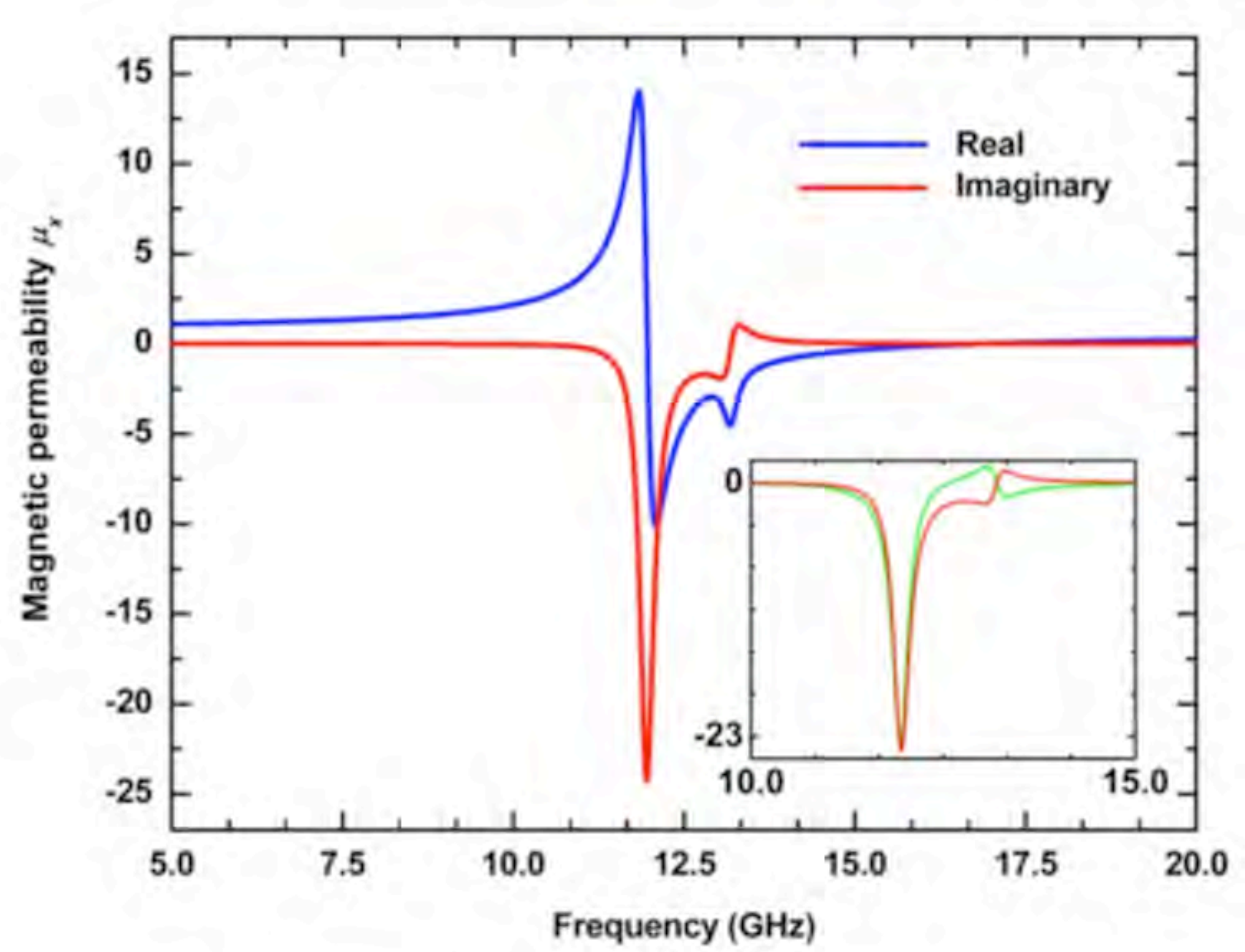

**Figure 3(b)**

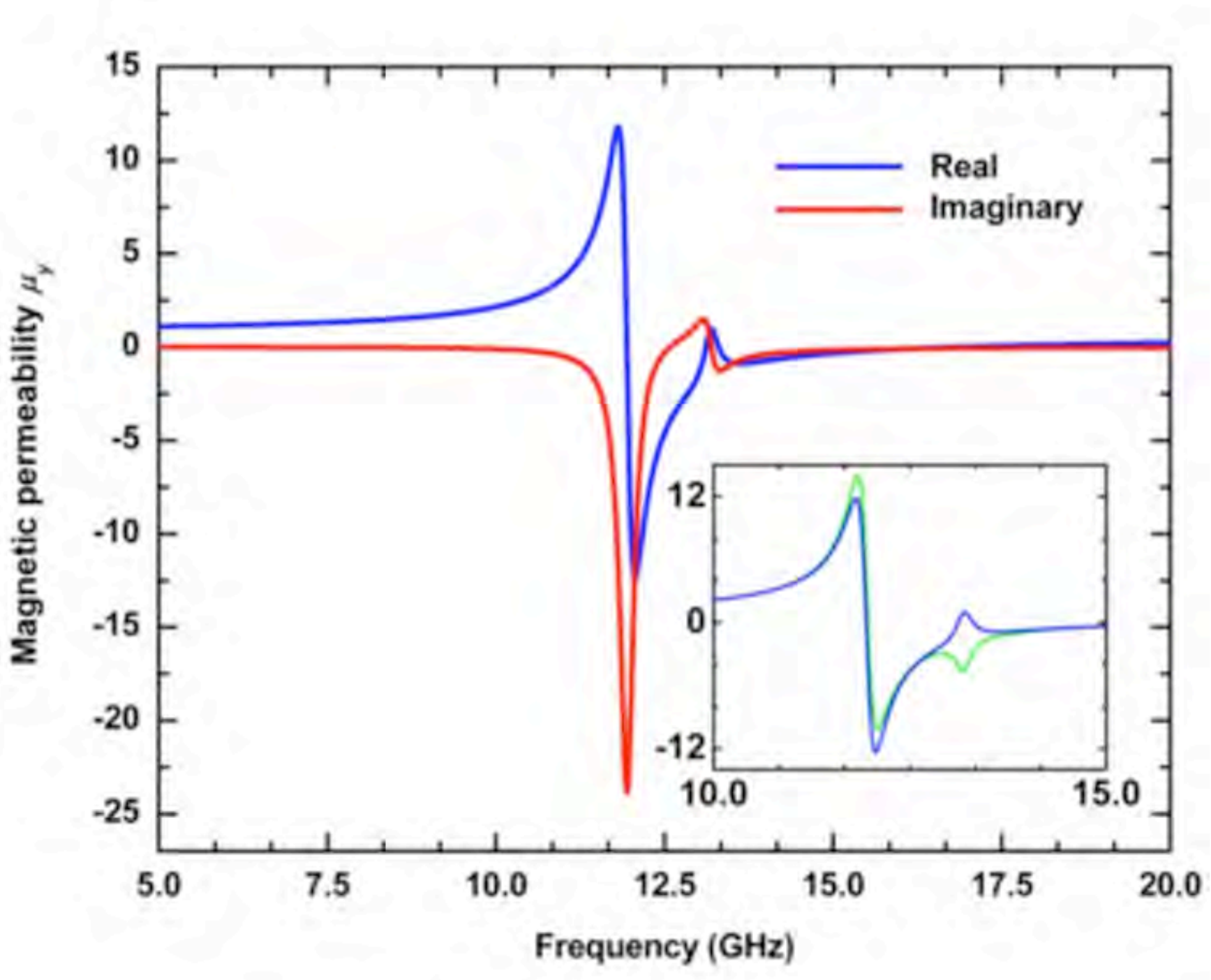

**Figure 3(c)**

# Supplementary Information

K. L. Tsakmakidis, M. S. Wartak, D. P. Aryal & O. Hess

# 1. Remarks on active and reactive powers

Consider two circuits, $C_1$ and $C_2$, electrically connected with two conducting wires, as shown in Fig. S1. Circuit $C_1$ contains an ideal source and passive elements (resistors, inductors, capacitors), while circuit $C_2$ contains only passive elements. Let us assume that the time-domain voltage, $V(t)$, and the current, $I(t)$, shown in Fig. S1, are sinusoidal functions of time, i.e.: $V(t) = V_0\cos(\omega t)$, and $I(\mathrm{t}) = I_0\cos(\omega t - \varphi)$, where $\varphi$ is the phase difference between the current and the voltage. For the sign convention shown in Fig. S1, a positive time-domain (instant) power, $P(t) = V(t)I(t) > 0$, means that *electrical power flows from circuit $C_1$ towards circuit $C_2$*. Conversely, if $P(t) < 0$ then *power flows from circuit $C_2$ to circuit $C_1$.*

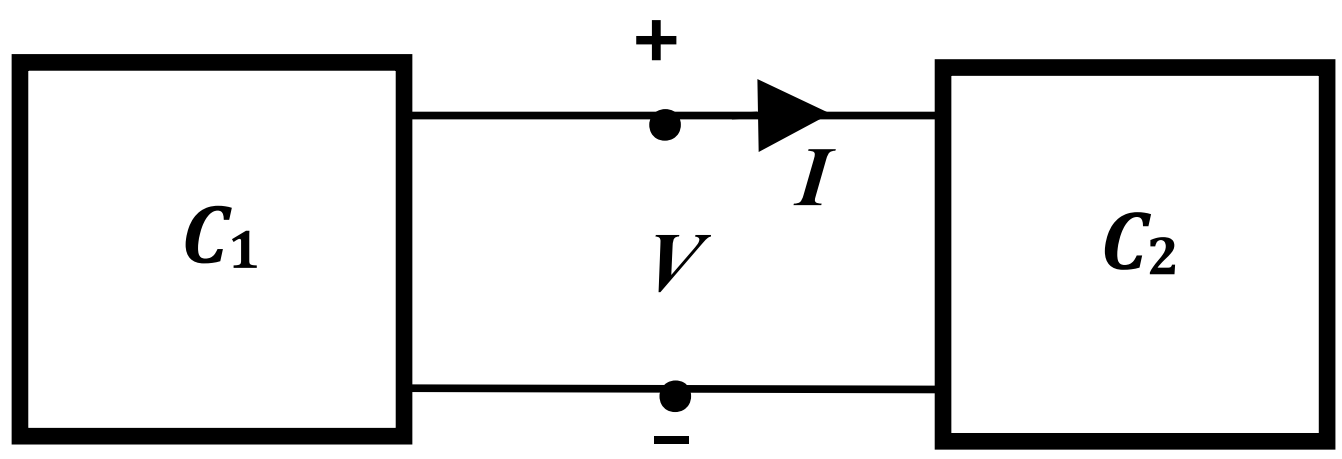

**Figure S1:** Schematic illustration of two electrically connected circuits.

If circuit $C_1$ contains only an (ideal) source, we always have: $P(t) > 0$ (power flows away from the source). It is, then, instructive to write the expression for the instant power $P(t)$ in the form:

$$P(t) = V_{\mathrm{rms}}I_{\mathrm{rms}}\cos\varphi[1 + \cos(2\omega t)] + V_{\mathrm{rms}}I_{\mathrm{rms}}\sin\varphi\sin(2\omega t), \qquad \text{(S1)}$$

where $V_{\mathrm{rms}} = V_0/\sqrt{2}$ and $I_{\mathrm{rms}} = I_0/\sqrt{2}$. Since we have assumed that the circuit $C_2$ contains only passive elements, the phase difference $\varphi$ between the current and the voltage will be: $-\pi/2 < \varphi < \pi/2$, and therefore the first term on the RHS of Eq. (S1) will be positive, corresponding to power flowing from circuit $C_1$ to circuit $C_2$, where it is consumed (dissipated). For this reason, this power, $P_R(t) = V_{\mathrm{rms}}I_{\mathrm{rms}}\cos\varphi[1 +$ $\cos(2\omega t)$], is called *real* or *active* power. The active power varies between 0 and $2V_{rms}I_{rms}\cos\varphi$ in one period, and a metric that is frequently used for it, is its time-averaged value: $\overline{P} = V_{rms}I_{rms}\cos\varphi$, which customarily is again referred to as *active* power. By contrast, the second term in the RHS of Eq. (S1), $P_x(t)$ = $V_{rms}I_{rms}(\sin\varphi)\sin(2\omega t)$, changes sign twice during the period $T = 2\pi/\omega$. During the first half-period, $P_x$ flows from $C_1$ to $C_2$, while in the second half-period the direction of the flow of $P_x$ is reversed, i.e. $P_x$ now flows from $C_2$ to $C_1$, so that the overall effect is zero. Because of the fact that $P_x$ does not (on average) produce any work, it is usually referred to as *reactive* power. Since the time-averaged value of $P_x$ is zero, a metric that is frequently used for its quantification is its *amplitude*, $Q = V_{rms}I_{rms}\sin\varphi$. Generally, the reactive power refers to the part of the instant power $P(t)$ that is sent by a source to the inductors and capacitors of the circuit, stored there temporarily, and then sent back to the source.

In many situations it is useful to consider the transformations of the voltage $V$ and current $I$ in the frequency domain, $\widetilde{V} = V_{rms}e^{i\theta_1}$ and $\widetilde{I} = I_{rms}e^{i\theta_2}$, $\theta_1$ and $\theta_2$ being the angles of the rotating vectors $\widetilde{V}$ and $\widetilde{I}$ with the real axis, and to work with the so called complex power, $S$:

$$S = \widetilde{V}\widetilde{I}^* = V_{rms}I_{rms}e^{i(\theta_1-\theta_2)} = V_{rms}I_{rms}e^{i\varphi}. \tag{S2}$$

It can, then, be readily observed that the reactive power is simply given by: $Q$ = $\mathrm{Im}\{S\}$, while the (time-averaged) active power is given by: $\overline{P} = \mathrm{Re}\{S\}$, which is the relation that we used in the calculations of the active powers referred to in the main text.

It should be noted that when the circuit $C_2$ does *not* contain sources (and $C_1$ contains only an ideal source), the active power $\overline{P_1}$ in $C_1$ is strictly positive, i.e. $C_1$ only *remits* real power to (does *not* receive real power from) $C_2$ wherein it is converted into heat at the resistances. In contrast, if $C_2$ also includes sources of electrical energy, then it is possible that $\overline{P_1}$ may become negative (for the sign convention shown in Fig. S1) in a frequency region, corresponding to active power being received by $C_1$, i.e. in that region $C_1$ acts as a *'load'*: it does not remit real power to $C_2$, but *receives* active power from the neighbouring meshes (in this case, $C_2$) that it is electrically connected with. However, the total active power, $P_{\text{tot}} = \sum_i P_i$ ($i = 1, \ldots, n$), $n$ being the total number of electrically connected meshes, must still be positive *at every frequency point*, since overall we have a net consumption of real power. These features are, indeed, precisely what we observed in the analytic calculations of the active powers that were reported in Figs. 2(c)–(e) in the main body of this work.

## 2. Calculations of effective magnetic permeabilities and active powers

In this section we shall obtain analytic expressions for the effective magnetic permeabilities of the configurations of Figs. 1(a) and (b), as well as for their associated active powers. Such expressions are useful because they elucidate various aspects of the magnetic responses of these structures.

A methodology for obtaining the effective permeabilities of configurations similar to those in Figs. 1(a) and (b) ('split ring resonators', SRRs) was first reported in [25].

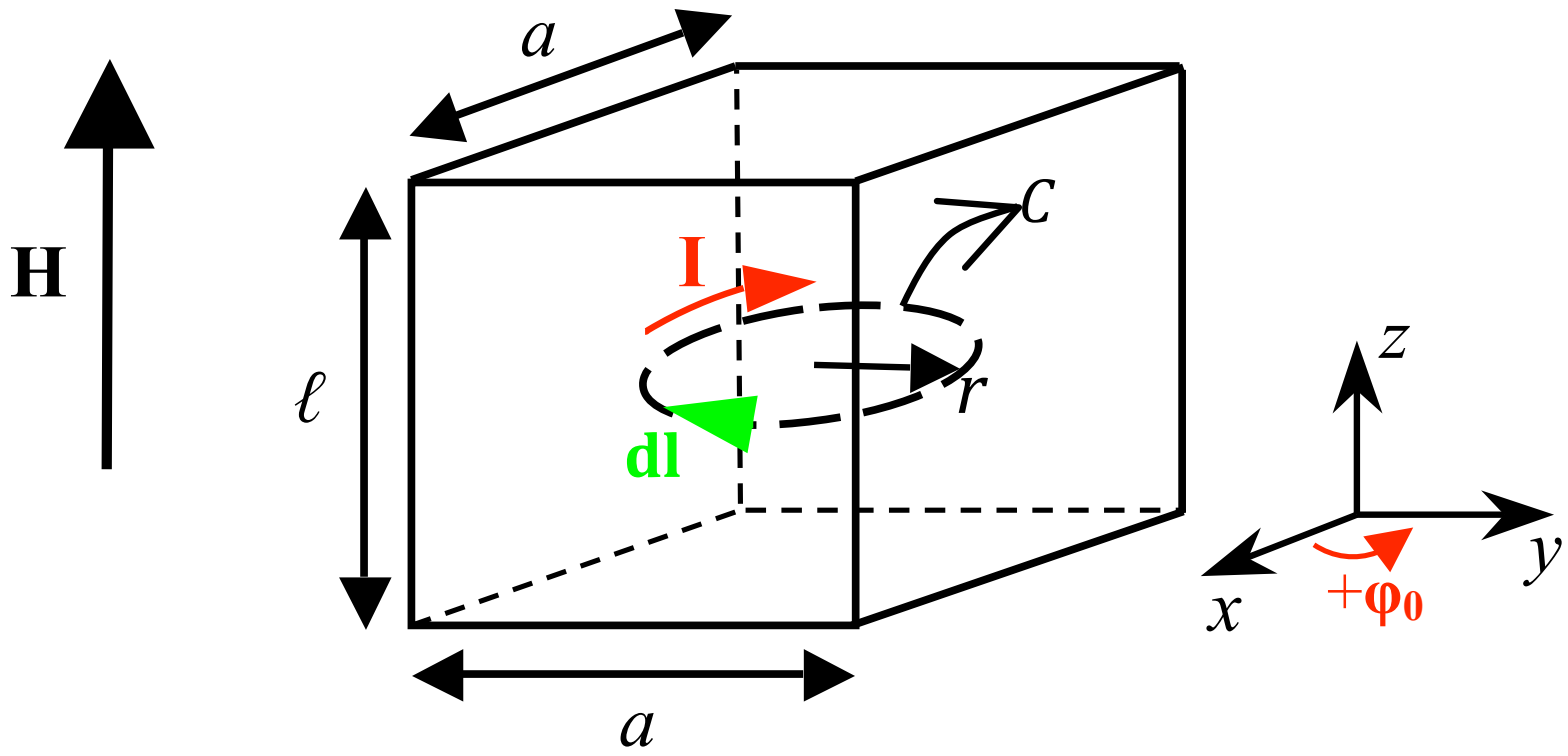

**Figure S2:** Unit cell containing a circular *RLC* metaparticle of radius $r$. The plane of the metaparticle is assumed to be parallel to the $xy$ plane. Also shown is the incident magnetic field, $\mathbf{H} = +H\mathbf{z_0}$.

In [S1]-[S3] similar expressions were obtained based on electric circuit equivalent models. Let us start by examining the *RLC* metaparticle of Fig. 1(a), which is assumed to be immersed in the parallelepiped unit cell of Fig. S2.

Application of Kirchhoff's second law to this circuit results in:

$$U_1 = \mathrm{abs}(U_{\mathrm{emf}}) = i\omega\mu_0 S H_0 = I[R + i\omega L + 1/(i\omega L)], \tag{S1}$$

where $\mu_0$ is the free space permeability and the remaining parameters are defined in the main article. From Lentz's law it follows that the current vector associated with the metaparticle will (with reference to Fig. S2) be: $\mathbf{I} = I(-\boldsymbol{\varphi}_0)$, while the induced magnetic moment will be: $\mathbf{m} = \frac{1}{2} I \oint_C \mathbf{r'} \times \mathbf{dl} = -I \int_S dS\mathbf{z}_0 = -I\pi r^2 \mathbf{z}_0$, with $\mathbf{r'}$ being a vector that begins from the origin of the coordinate system and ends at the element **dl**. Recalling that the self-inductance of a 'pile' of such 'SRRs' will (in the solenoid approximation) be $L = \mu_0 S/\ell$, $\ell$ being the vertical distance between SRRs in the pile, the magnetisation $\mathbf{M} = \mathbf{m}/G$ of a structure comprised of a 3D periodic arrangement of the SRRs, with a parallelepiped unit cell of volume $G = a^2 \times \ell$, will be:

$$\mathbf{M} = FH_0 / [1/(\omega^2 LC) - 1 + iR/\omega L]\mathbf{z}_0, \tag{S2}$$

with $F = S/a^2$ being the unit cell filling factor. Assuming low-density, weakly interacting ('isolated') unit cells and that the incident wavelength $\lambda >> \max\{a, \ell\}$ (quasi-static approximation, which accurately describes the behaviour of an SRR around its *first* resonance; see, e.g., [S4]), results in the following expression for the (relative) effective permeability of the structure:

$$\mu = 1 + M / H_0 = 1 + F / [1/(\omega^2 LC) - 1 + iR/\omega L]. \tag{S3}$$

From Eq. (S3) one immediately sees that in this case we have: $\mathrm{Im}\{\mu\} < 0$, i.e., for the considered $\exp(i\omega t - i\mathbf{k}\bullet\mathbf{r})$ spatiotemporal dependence, the structure is magnetically *passive*, as expected.

Let us, now, consider the configuration of Fig. 1(b) where an external ac source $U_2' = iU_2$ is electrically connected to the *RLC* metaparticle *within each unit cell*. The metaparticle is, again, assumed to be immersed in the parallelepiped unit cell of Fig. S2. In this configuration we have: $|U_2'| > |U_1|$, i.e. $U_2 > \omega\mu_0 SH_0$, and the $U_2'$ source has opposite polarity compared to the electromotive source $U_1$. Here, one may either directly apply Kirchhoff's law of voltages, as in the previous example, or one may separately calculate the vector currents associated with each source and then add them (since the circuit is linear) in order to find the total current $I$. Following the second methodology, we obtain: $\mathbf{I_1} = -(U_1/Z)\boldsymbol{\varphi}_\mathbf{0}$, $\mathbf{I_2} = +(iU_2/Z)\boldsymbol{\varphi}_\mathbf{0}$, with $Z = R + i\omega L - i/(\omega C)$, from where the total current is calculated as: $\mathbf{I} = \mathbf{I_1} + \mathbf{I_2} = \{(iU_2 - i\omega\mu_0 SH_0)/Z\}\boldsymbol{\varphi}_\mathbf{0}$. We note that since $U_2 > \omega\mu_0 SH_0$, the current $\mathbf{I}$ will be circulating in the *RLC* metaparticle along the $+\boldsymbol{\varphi}_\mathbf{0}$ direction (see also the expressions for the active powers below). It follows that the magnetization $\mathbf{M}$ of the structure will be directed along the $+\mathbf{z_0}$ direction, i.e. $\mathbf{M} = +M\mathbf{z_0}$, in the same direction as the incident magnetic field (thereby reinforcing/amplifying it). Indeed, the effective (relative) permeability of this structure is given by:

$$\mu = 1 + \frac{F}{\ell H_0}\frac{(U_2 - \omega\mu_0 SH_0)(W + iR)}{R^2 + W^2}, \tag{S4}$$

with $W = \omega L - 1/(\omega C)$, from where one can see that $\mathrm{Im}\{\mu\} > 0$ (magnetically *active*).

For the configuration of Fig. 1(b), where the current has been reversed and the induced magnetisation $\mathbf{M}$ is co-directed to the incident magnetic field, one can also straightforwardly calculate the active power emitted by the source $U_2'$ and consumed at its 'load' (resistance $R$ and potential barrier of the emf source $U_{\mathrm{emf}}$). The active power emitted by the source then is:

$$\mathrm{Re}\{U_2^{/}I^*\} = \frac{U_2(U_2 - \omega\mu_0 S H_0)}{R^2 + W^2} R > 0, \quad \text{(S5)}$$

while the active power consumed at the resistance $R$ and in overcoming the potential barrier of the electromotive source $U_1$ are, respectively:

$$|I|^2 R = \frac{(U_2 - \omega\mu_0 S H_0)^2}{R^2 + W^2} R > 0, \quad \text{(S6)}$$

$$\mathrm{Re}\{U_1 I^*\} = \frac{\omega\mu_0 S H_0 (U_2 - \omega\mu_0 S H_0)}{R^2 + W^2} R > 0. \quad \text{(S7)}$$

We note from Eqs. (S5)-(S7) that: $\mathrm{Re}\{U_2^{/}I^*\} = |I|^2 R + \mathrm{Re}\{U_1 I^*\}$, i.e. the active power is, indeed, conserved in this system. Similar relations hold for the configuration of Fig. 1(a).

# 3. Remarks and discussion

The presented two-degrees-of-freedom based (2-DEG; see section-7 herein) methodology allows for the realisation of ultralow- (Fig. 2) or lossless/active (Fig. 3) magnetic metamaterials over a continuous range of frequencies. As highlighted in the main text, a number of approaches have previously been proposed for compensating losses [S9], [S10] or even creating stable gain [S11] in metamaterials. Those approaches usually relied on providing optical gain, e.g. in the form of optical parametric amplification [S9] or electromagnetically induced chirality [S10], or on the use of active, negative-resistance, diode elements, such as Gunn or resonant tunnel diodes [S11], which enable 'cancelling' the ohmic losses of the metaparticles and possibly even lead to magnetic metamaterials with gain. However, though promising, such approaches normally require high field intensities and/or are accompanied by nonlinearities in the response of the engineered medium, which may not always be desirable. Moreover, their scaling from radio to visible frequencies (or vice versa) could be challenging. Clearly, a much more desirable and convenient approach would be to judiciously redesign the structure of the metaparticles at the unit-cell level, such that it could open a 'window' for harnessing perfectly lossless artificial magnetism over a continuous range of frequencies. This is the route followed here.

The approach presented herein is based solely on an exchange of active powers between the two electrically (capacitively) connected *RLC* metaparticles inside each unit cell. It should be emphasized that the exchange of active power between the two metaparticles occurs to such an extent that, crucially, *it enables reversal* of the direction of the current circulating in one of the metaparticles *without affecting the magnetic field perpendicular to it*. It follows that the magnetic field incident on that metaparticle can profit (be amplified) from the induced magnetisation, which is now co-directed to the **H**-field incident on that metaparticle. Indeed, there are, as expected, frequency regions wherein the imaginary parts of the susceptibilities associated with the two meshes become positive (but not simultaneously at the same frequency region), signifying active metamaterial magnetism.

It should also be noted that such an exchange of active power between two (physically separated) light beams *does not occur in any conventional linear medium* (in which light beams *can only exchange power with the aid of nonlinearity*). In our scheme, this coupling of the two orthogonal light beams occurs owing to the special design of the structure of the meta-molecules – which are electrically coupled and can, thereby, facilitate 'transfer' of active power from one mesh to the other. Such an exchange of active power *can*, in general (as our analysis has also shown), occur even between two *strictly linear*, electrically connected meshes. In our structure, the aforementioned active power exchange further requires that the ohmic resistances of the two meta-molecules are substantially different (e.g., $R_1 >> R_2$ or vice versa), assuming that $L_1 = L_2$ and $C_1 = C_2$ (see also below). Since this type of amplification depends on a suitable choice of optogeometric parameters (that determine the $R$, $L$, $C$ values) and can *only* occur in specially designed metamaterials (*sui generis*), we refer to it as *‘metamaterial parametric amplification’* (MPA) – not to be confused with optical parametric amplification (OPA) that can also occur in metamaterials [20], [21].

It is further to be noted that with the present anisotropic design we are able to harness zero-loss negative-$\mu$ magnetic metamaterials – indeed, even with artificial magnetic gain – at the cost of having *increased magnetic losses in the other direction*, as well as heat generated along the direction in which the magnetic field is amplified. For instance, assuming that the imaginary part of $\mu_1$ is *positive* in a frequency region, one will observe an *increase* in the magnetic density $B_y$ in the direction of its propagation, owing to $\mathrm{Im}\{\mu_1\} > 0$; at the same time, however, the *decrease* (owing to $\mathrm{Im}\{\mu_2\} < 0$ in the same frequency region) of the magnetic density $B_x$ along the direction in which it propagates will be *larger* compared to the increase that $B_y$ experienced (Figs. 3(b) and (c)), so that – at every frequency point – there will be no violation of the conservation of the magnetic energy, since for every frequency we have: $\mathrm{Im}\{\mu_1\} + \mathrm{Im}\{\mu_2\} < 0$ (no ‘net’ gain, i.e. there cannot be gain generated, simultaneously, at both directions *and* at the same frequency region). The reason that, in this example, the magnetic losses associated with the second mesh increase is that, in order for the voltage $V_{ab}$ (Fig. 1(c)) to become larger than $V_1$ (and thereby reverse the current $I_1$ in the first mesh), the current $I_2$ of the second mesh has to resonantly increase – a process that unavoidably leads to increased dissipative losses associated with the second mesh.

A somewhat similar situation wherein *increased* dissipative losses lead to field *amplification* has also recently been noticed in schemes deploying ‘conventional’ OPA to overcome losses in metamaterials [20]-[21]. Similarly to those schemes, the configurations of Figs. 1(b) and (c) *require* resistive losses to accomplish amplification of the magnetic field. Indeed, inspection of Eq. (S4) reveals that for $R = 0$ it is $\mathrm{Im}\{\mu\} = 0$. Physically, this result is justified from the fact that when $R = 0$ there is, over the first half cycle of the oscillation, power *sent* by the sources to the inductors and capacitors of the system, but this power is again *sent back* to the sources over the next half cycle, i.e. the overall effect to the incident magnetic field of this *reactive* power is zero (see also section-1 herein). Indeed, as one may see from Eqs. (S4)-(S6), for $R = 0$ the active powers associated with the two sources vanish (and, of course, so does the active power consumed at the resistance $R$), and as a result, in this case, active power exchange between the two meshes cannot occur.

It is interesting to note that frequency regions with $\mathrm{Im}\{\mu\} > 0$ are also frequently encountered in numerical or experimental calculations/extractions of effective electromagnetic parameters (refractive index $n$, permittivity $\varepsilon$ and permeability $\mu$) of composite metamaterial structures [S5]-[S10]. Unfortunately, in those cases the frequency regions where $\mathrm{Im}\{\mu\} > 0$ always correspond to incident wavelengths that are equal to or smaller than the periodicity of the structure, i.e. the calculated permeabilities (or permittivities) are *pseudo-effective*; they are only an artefact of the periodicity of the structure and always disappear when the incident wavelength becomes considerably larger than the periodicity [S11]. Moreover, in those cases there is always a strong coupling between the magnetic and electric responses of the composite medium, as a result of which at a frequency region where $\mathrm{Im}\{\mu\} > 0$, it is also that: $\mathrm{Im}\{\varepsilon\} < 0$ (resonance-antiresonance effect), so the overall electromagnetic response of the medium is still passive ($\mathrm{Im}\{n\} < 0$). By contrast, in the herein presented scheme, the frequency regions where $\mathrm{Im}\{\mu\} > 0$ occur because, as was explained above, one mesh is able to 'pump' active power to the second mesh and reverse the direction of the current circulating in the second mesh. The reversal in the sign of $\mathrm{Im}\{\mu\}$ occurs solely because of the aforementioned effect, i.e. it has an 'all-magnetic' origin and does not occur because of a magneto-electric coupling. In fact, as was pointed out in the main text, our structure can be designed such that it is non-bianisotropic (see also section-5 herein), either by, e.g., ensuring that the plates of the lumped capacitors are not perpendicular (i.e. they do not couple) to the incident electric field or by deploying physical, *non-bianisotropic* split-ring resonators [23], [S4], [S12]-[S13], etc. Such a structure will be characterised by an effective $\varepsilon \approx 1$ ($\mathrm{Im}\{\varepsilon\} = 0$) for all frequencies of operation, i.e. it will be a purely magnetic metamaterial.

Our structure can, also, potentially be *electromagnetically active*, as long as it is combined with a suitable (active *or even passive*) purely-dielectric metamaterial. This result will not, of course, violate energy conservation since the amplification of a light beam will occur (in a given frequency region) only along *one* direction, enabled by the additional energy that will be 'pumped' by a second light beam propagating at an orthogonal direction to the first beam – similarly (but in a completely linear fashion) to two-beam coupling 'gain' or optical parametric amplification. Note that, as was pointed out above, this second light beam will itself be experiencing *increased* (compared to the case where the first beam was absent) dissipative losses along the direction that it propagates inside the metamaterial (see Figs. 3(b) and (c)).

From the examples presented in the main text it should also be clear that the present scheme does not require large intensities for the incident fields – in fact, it works equally well even with low field intensities. Indeed, all the results shown in the main body of this work were obtained assuming incident magnetic fields having intensity of just 1A/m. Obviously, similar results can be obtained with even smaller magnetic intensities, so long as they suffice to excite a useful or interesting collective magnetic response of the effective medium.

The approach introduced here for creating lossless or active magnetic metamaterials can be most conveniently realized experimentally in the radio and microwave frequencies by using discrete lumped resistors, inductors and capacitors, or non-bianisotropic SRRs placed at different planes and made of different conductors. The scheme is also scalable down to optical frequencies (see also section-4 herein), where magnetic metamaterials made of arrays of SRRs have already been demonstrated. A further method for the construction of the herein proposed structures at optical frequencies could be the use of discrete nanoresistors, nanoinductors and nanocapacitors [S14], which have already been studied and were shown to hold promise in connection with the creation of optical nanocircuits [S15] and nanoantennas [S16].

Finally, it should be noted that the electrical (capacitive) connection of the meshes is a crucial aspect of the proposed mechanism for overcoming metamaterial losses, not only because it allows electrical power to flow and be exchanged more easily between the meshes, but also because it allows each magnetic field component to generate a magnetic moment at both, the plane to which it is perpendicular *and* at a plane to which it is parallel. For instance, the $H_y$-field component generates currents circulating in *both* meshes residing at the *xz* and *yz* planes. As a result, the $H_y$-field component induces a magnetic moment not only along the *y*-direction (perpendicularly to the *xz* plane), but also along the *x*-direction (perpendicularly to the *yz* plane, to which the $H_y$-field component is parallel). The corresponding is, of course, also true for the $H_x$-field component. This superposition of the magnetic moments generated, at both planes, by both **H**-field components is an essential feature of the present design. Note also that in order for the 'superposition principle' to hold our present scheme *requires* linearity and low or moderate field intensities to function according to its conception.

## 4. Robustness to dissipative losses and extension to the optical regime

It turns out that the underlying mechanism (i.e., the exchange of active power between the meshes) that is responsible for enabling lossless/active metamaterial magnetism is extremely resilient and robust to the presence of ohmic losses, insofar as the *difference* in (*not the actual values of*) the ohmic resistances of the meshes is substantial – in the examples of the main text it was: $R_1 = 50\ \Omega$ and $R_2 = 0.1\ \Omega$.

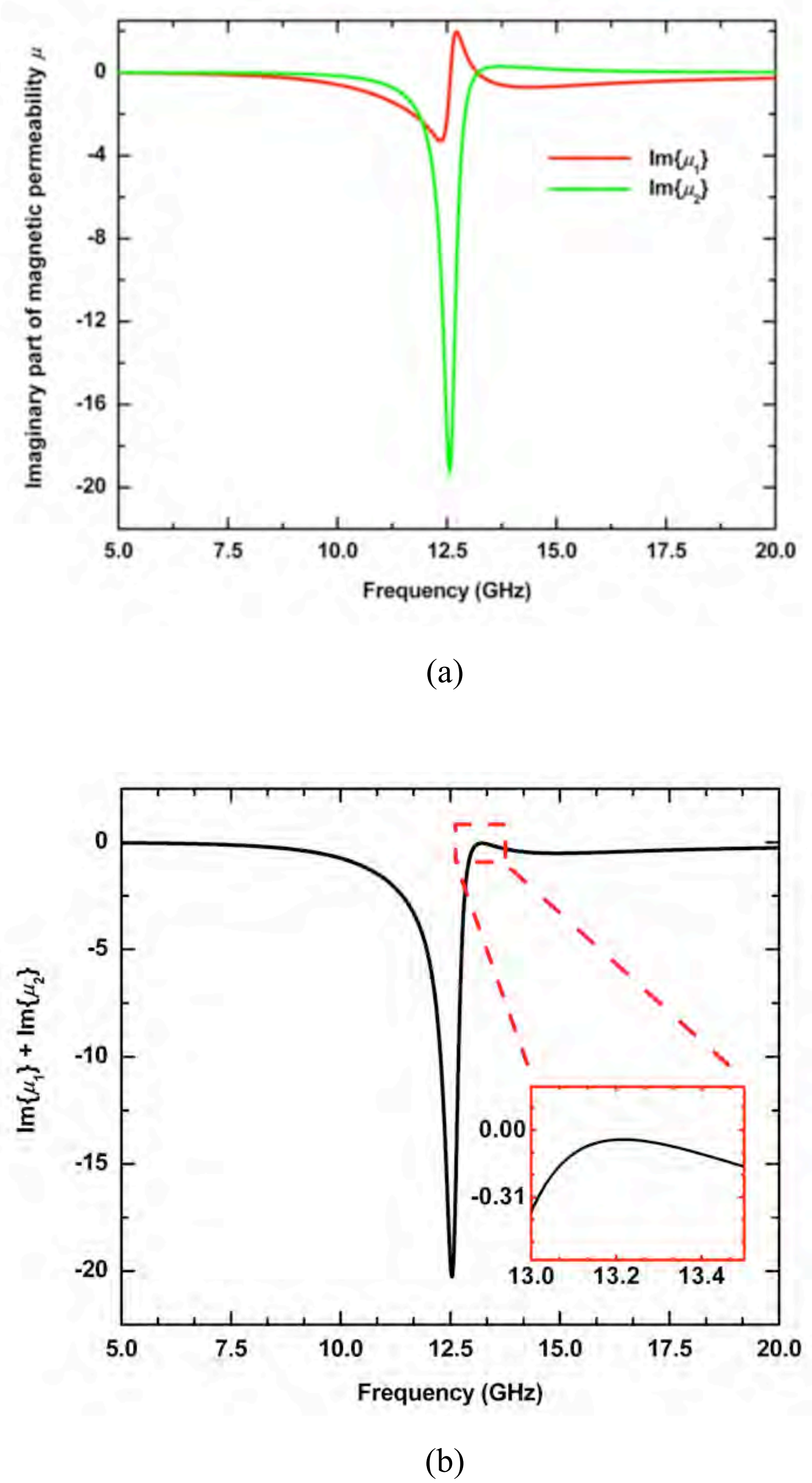

**Figure S3**: (a) Imaginary parts of $\mu_x$ and $\mu_y$ for the electromagnetic and geometric parameters used in Fig. 3, but now with $R_1$ = 500 Ω and $R_2$ = 1 Ω. (b) Sum of the imaginary parts of $\mu_x$ and $\mu_y$.

Indeed, from Fig. S3 one observes that even when the resistance in each mesh increases by an order of magnitude ($R_1$ = 500 Ω, $R_2$ = 1 Ω) there are, as before, regions wherein the imaginary parts of the effective permeabilities become positive (and with: $\mathrm{Im}\{\mu_1\} + \mathrm{Im}\{\mu_2\} < 0$ for every frequency; see Fig. S3(b)). The same holds true for any increase in $R_1$ and $R_2$, to the extent that we do not enter the 'overdamped' region wherein there is no effective magnetic oscillation (response) in the effective medium at all. Ultimately, this robustness against losses is, as highlighted before, owing to the fact that the underlying mechanism relies critically on an *imbalance* in the values of the resistances of the pair of meshes, and *not* on the actual values of the resistances themselves. As long as such an imbalance is present (and provided that we are not in the 'overdamped' regime) there will always be power flowing *away* from a mesh to its partner and reversing the current circulating in this second mesh. This will result in algebraically negative active power associated with the source in the electrical mesh whose current has been reversed, and in positive imaginary part for the corresponding magnetic susceptibility/permeability.

Remarkably, considering the fact that typical values of ohmic resistances in the radio and microwave regime are usually less than 1 Ω, our above results demonstrate that even when we use a resistance $R_1$ for an individual metaparticle that is *more than four orders of magnitude* larger compared with a typical value (e.g., from a typical value of $R_1$ < 1 Ω, to a value $R_1$ = 500 Ω used in Fig. S3) there still exists a continuous frequency region with $\mathrm{Im}\{\mu_1\} > 0$.

Furthermore, all the presented results are scalable and still hold unchanged *even in the optical regime*. Indeed, a typical value for the ohmic resistance of an individual metaparticle in the optical (infrared @ $\lambda$ = 1.4 μm, $f \approx$ 214 THz) regime is $R \approx 5$ Ω [24] – the radiation resistance [S17] for an assumed *deep-subwavelength* metamaterial structure will, of course, be zero; absence of radiation effects and/or sensitivity on, e.g., the phase difference between the two waves/beams, is assured by suitably constructing the magnetic metamaterial such that it is, e.g., electrically small. Thus, for a corresponding calculation in the infrared regime, we may use the previous values of $R_1$ = 500 Ω and $R_2$ = 1 Ω, and scale all the geometric and electric-circuit parameters (apart from the resistances) by, e.g., a factor of ≈ 17,000. A calculation (not shown here for the sake of brevity) of the new effective permeabilities $\mu_1$ and $\mu_2$ reveals that their variation with frequency is exactly the same with the corresponding ones of Fig. S3, the only difference being that the new resonance frequency will now be: $f_0 \approx 225$ THz. Thus, there are again frequency regions where $\mathrm{Im}\{\mu_1\} > 0$ or $\mathrm{Im}\{\mu_2\} > 0$. This 'scalability' of the $\mathrm{Im}\{\mu\}$ with frequency (assuming the *strict* condition that the value of the resistance $R$ is *constant and equal to that occurring at the higher frequency regime*) can also be directly inferred from Eq. (S4) assuming that $U_2 \propto \omega$. Furthermore, it turns out that the present magnetic metamaterial structure(s) can readily be designed such that a *purely real* (and negative) magnetic permeability can be achieved (e.g., $\mu = -1 + i0$, for 'perfect lenses' or light-stopping metamaterial structures) simply by, e.g., increasing the electrical size of the metamaterial structure, such that for phase differences between the two beams $\Delta\varphi \in [0^o, 180^0]$ it will, e.g., be: $\mathrm{Im}\{\mu_x\} < 0$, while for $\Delta\varphi \in (181^o, 359^0)$ it will be: $\mathrm{Im}\{\mu_x\} > 0$, with the overall $\mathrm{Im}\{\mu_x\} \approx 0$.

The above results suggest the remarkable (and counterintuitive) prospect of actually deploying 'poor' conductors to create a perfectly lossless magnetic metamaterial in a specified direction (or directions; see section-7 herein). For instance, let us assume that an 1-DEG system, corresponding to the equivalent electrical circuit of a split ring resonator (SRR), uses 'good' conductors of resistance 0.1-1 Ω, which is typical for conductors in the GHz regime [25]. It is well-known that a periodic arrangement of the single *RLC* mesh in this 1-DEG system will, in the quasi-static regime, result in an effective medium exhibiting artificial magnetism [25], but limited by the presence of the ohmic losses.

Instead of attempting to further reduce the losses present in this medium (by, e.g., using even better conductors for the single meshes), our analysis shows that one may (or, in fact, should) deploy a ‘poor’ conductor of resistance 50-500 Ω to create a second *RLC* mesh on a different plane than the first. The imbalance in the values of the resistances of the two meshes residing at the two planes will give rise to power being exchanged between the two meshes which, according to what was explained above, will cause positive imaginary parts for the associated permeabilities over certain frequency regions, and therefore in perfectly lossless magnetism in the corresponding directions – in fact, even with the presence of gain.

## 5. Non-bianisotropic design of the 2-DEG magnetic meta-material

In the schematic illustration below we are showing an example of how the structure studied in Fig. 3(a) can, most straightforwardly, be re-designed such that it is fully non-bianisotropic, while maintaining the magnetic response of the structure shown in Fig. 3(a). Indeed, for $C_1^{/}=2C_1$, $C_2^{/}=2C_2$ and $C^{/}=2C$, the magnetic response of this structure will be *exactly* the same with that of the structure of Fig. 3(a) in the main text, but now this new structure (shown below) will be characterized by an effective electric permittivity $\varepsilon = 1 + i0$ in all directions (electrically isotropic).

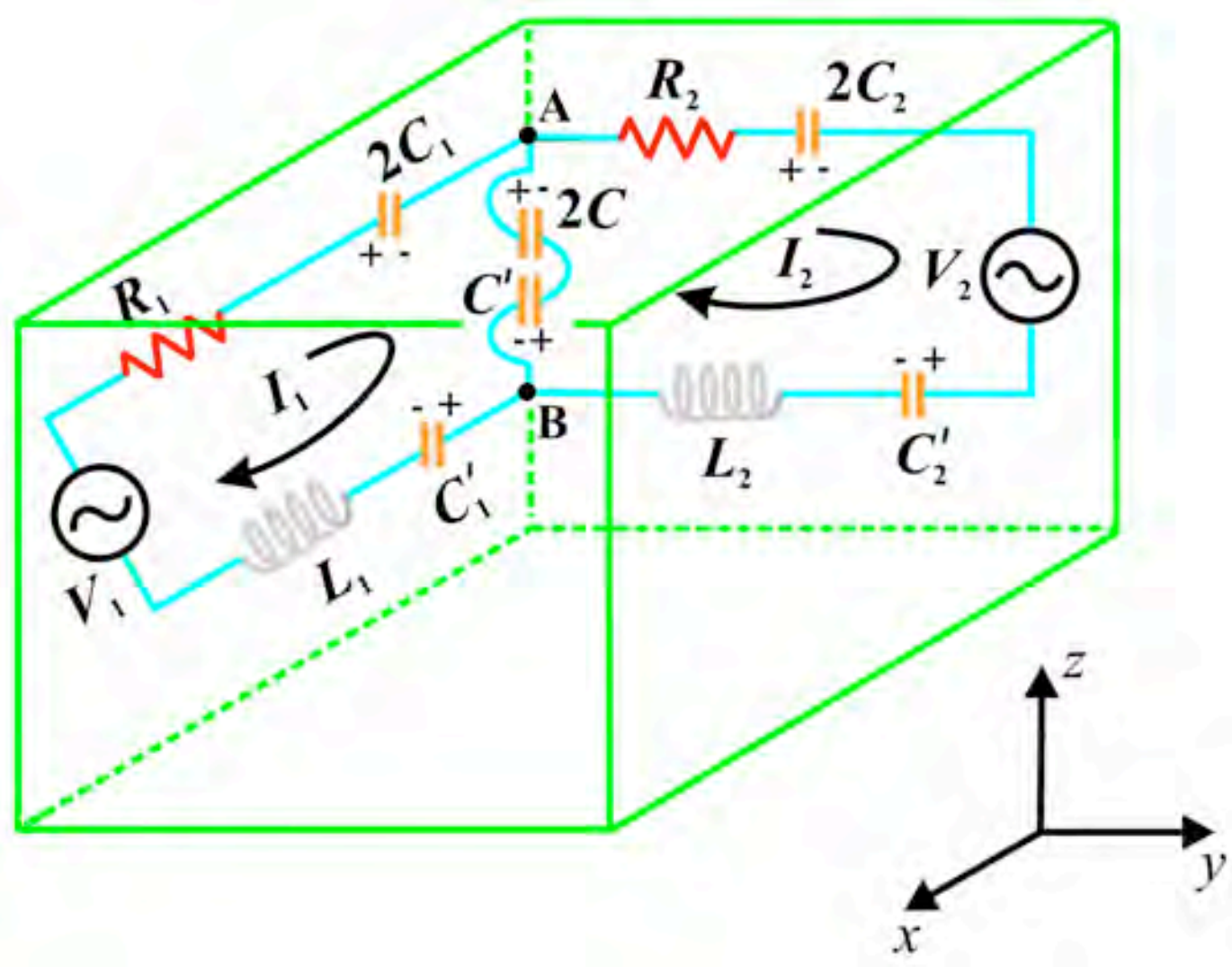

**Figure S4**: Non-bianisotropic implementation of the configuration shown in Fig. 3(a).

In the above structure it is assumed that the electric field for both incident light beams is polarized along the $z$-direction (parallel to the plates of all capacitors), i.e. it does not couple to any capacitor. Moreover, following the methodology for the design of non-bianisotropic magnetic metaparticles found in [23], [S4], [S12], [S13], one can readily devise a scheme (such as the one shown above) that can enable *cancelling/'neutralizing'* the dipole moments generated by the magnetic fields in the capacitors of the structure of Fig. 3(a). Here, the charge accumulated in the capacitor $2C_1$ is equal to the charge accumulated to capacitor $C_1^{/}$ (these two capacitors are connected in series), but the dipole moments generated in these two capacitors are *oppositely* directed (and of the same magnitude if the separation of the capacitors' plates is the same for both capacitors); hence, they infallibly cancel each other. The same, of course, also holds true for the capacitor pairs {$2C_2$ and $C_2^{/}$} and {$2C$ and $C^{/}$} [in the Fig. S4 above we have assumed that $I_1 > I_2$, so that current (of positive carriers) flows downwards, from point A to point B; if $I_1 < I_2$, then the 'polarity' of the capacitors $2C$ and $C^{/}$ will be reversed, without affecting any of the conclusions mentioned above]. Note, also, that in the structure of Fig. S4, even the dipole moments generated (along diagonal directions) by neighbouring capacitors (e.g., $C_1^{/}$ and $C^{/}$, etc) mutually cancel. This structure is, thus, in every sense a fully non-bianisotropic one, characterized by an effective permittivity $\varepsilon = 1 + i0$ in all directions.

It is to be noted that, as we have pointed out on the main text, the summation of the imaginary parts of $\mu_x$ ($\mu_2$) and $\mu_y$ ($\mu_1$) for the structure shown in Fig. 3(a) is, at every frequency, exactly equal to the imaginary part of $\mu$ shown in Fig. 2(a) (i.e.: $\mathrm{Im}\{\mu_x\} + \mathrm{Im}\{\mu_y\} = \mathrm{Im}\{\mu\}$). Thus, the summation $\mathrm{Im}\{\mu_x\} + \mathrm{Im}\{\mu_y\}$ for the structure of Fig. 3(a) is, for every frequency, 'locked' to the corresponding value of $\mathrm{Im}\{\mu\}$ that occurs when both meshes are on the same plane (Fig. 1(c)). As a result, when e.g. the term $\mathrm{Im}\{\mu_x\}$ reduces abruptly towards zero (signifying reduced magnetic losses for an $x$-polarised magnetic field component) or changes sign (signifying magnetic gain for an $x$-polarised magnetic field component) this reduction or reversal in sign is correspondingly 'absorbed' by the term Im$\{\mu_y\}$ – so that their summation remains constant and equal to Im$\{\mu\}$. It follows that the dissipative losses that, e.g., an $x$-polarised magnetic field avoids are 'transferred'/added to the dissipative losses that the $y$-polarised *magnetic* field experiences – i.e., they are *not* 'transferred' (via, e.g. a magneto-electric coupling, etc) to an assumed *electric* response of the structure (such an electric response ($\varepsilon \neq 1$) of the structure is or can be, as we explained above, infallibly absent). Thus, the mechanism behind the ultralow-loss or active magnetic behaviour of our structure has an *all-magnetic nature*, where dissipative losses are 'transferred' from the *magnetic* response of the structure along one direction to (*solely*) the *magnetic* response of the structure along another (orthogonal to the first) direction.

## 6. Effective magnetic permeabilities of 2-DEG meta-materials for the case of 'tightly coupled' unit cells

In the main text of the article we studied the case where the metaparticles at the centres of neighbouring unit cells are sufficiently apart from each other, i.e. the filling factor – defined as the ratio between the total area of the two meshes and the area of the basis of a unit cell [S2]-[S3] – is sufficiently small and the cells can then be regarded as weakly interacting ('isolated'). This is typical of low-density (dilute) gases, but note that contrary to the case of gases wherein the weak interaction of the molecules leads to the absence of any magnetic response and to $n \approx 1$, metamaterials can be designed to exhibit strong magnetic response [25], even for low densities, simply by increasing the $Q$-factor (e.g., by decreasing the value of the resistance $R$) of the resonant $RLC$ meshes. Under these conditions, higher-order multipole terms are, indeed, negligible [S19] and the local fields in each cube are just $H_{loc} = H_0$ and $B_{loc} = B_0$, where $B$ is the magnetic flux density, so that we may define $\mu_{r,eff} = 1 + M/H_0$.

Note, also, that in this case the effective (relative) permeability is directly proportional to the magnetization $M$.

If the metaparticles at neighbouring unit cells are 'tightly coupled', then a possible expression for $\mu_{r,eff}$ can be [S1]:

$$\mu_{r,eff} = \frac{1 + \frac{2}{3}\frac{M}{H_0}}{1 - \frac{1}{3}\frac{M}{H_0}} . \tag{S8}$$

In this case, $\mu_{r,eff}$ is not any more simply proportional to neither the magnetisation $M$ nor to the *total* active power $P = P_1 + P_2$, $P_m = \mathrm{Re}\{V_m I_m^*\}$ ($m = 1, 2$) of the pair of meshes, and the reported effects become slightly more intricate, but do not fundamentally nor qualitatively differ from the results discussed so far. It is useful to also note that in the 'tight coupling' case, and for meshes that are stacked up closely together (solenoid approximation), the uniform depolarization magnetic field [25] only modifies the value of the inductances $L_m$ in each cube [S3], i.e. merely reducing the value of $L_m$ in each electric mesh is sufficient to incorporate the effect of the depolarisation field in the analysis. Thus, in order to facilitate direct comparison with the corresponding results that were presented so far, one can retain the same value (16 nH) for the inductance of each mesh as in the main body of the article. This is, of course, equivalent to assuming that the actual inductance $L$ in each mesh is slightly larger, so that when $M$ is subtracted from $L$ we end up with a total inductance in each mesh equal to 16 nH.

Pursuing such an analysis (the results of which are not shown here for brevity) reveals that the variations with frequency of the active powers in each mesh are, as expected, precisely the same as the corresponding ones presented in the main article – all the electric circuit equations remain unchanged. Accordingly, all the dips or peaks in the imaginary part of the effective permeabilities associated with the exchange of active power between the meshes continue to occur at the same frequency (≈ 12.5

GHz) as in the case of 'isolated' cells. However, since the effects of neighbouring cells are now incorporated in the analysis, one does expect to observe a redshift in the collective resonant response of the composite medium. For instance, it is well-known from the classical Lorentz theory for the description of dielectric molecules that the use of the Clausius-Mossotti relation (which is very similar to Eq. (S8) above) results in a redshift of the resonant frequency from $\omega_0^2$, which occurs the density of the molecules is reasonably low ('isolated' unit cells, as in a gas), to $\omega_0^2 - Ne^2/3\varepsilon_0 m_e$, where $N$ is the density of the molecules, $e$ the electronic charge, $m_e$ its mass, and $\varepsilon_0$ the free-space permittivity. This is also what we have observed in the new calculations.

Interestingly, though, we also noticed that using Eq. (S8) to describe the quasi-static response of a magnetic metamaterial composed of 'tightly coupled' meshes at neighbouring unit cells, resulted in a frequency region where the summation of the imaginary parts of the effective permeabilities in the two orthogonal directions became positive ($\mathrm{Im}\{\mu_1\} + \mathrm{Im}\{\mu_2\} > 0$), implying the occurrence of 'net' gain in that region. As explained in the following, such an outcome emerges ultimately owing to the violation of the assumptions used in the derivation of Eq. (S8).

Indeed, let us start by noting that frequency regions wherein $\mathrm{Im}\{\mu_1\} + \mathrm{Im}\{\mu_2\} > 0$ *never* occur for the case of low-density, weakly interacting ('isolated') unit cells. This is simply because, in this case, the sum $\mathrm{Im}\{\mu_1\} + \mathrm{Im}\{\mu_2\}$ is directly proportional to the *total* magnetisation $M$ of the structure and, thus, ultimately, directly proportional to the *total* (cycle averaged) active power emitted by the two emf sources, $P = P_1 + P_2 > 0$, which only assumes positive values at every frequency. This proportionality of the term ($\mathrm{Im}\{\mu_1\} + \mathrm{Im}\{\mu_2\}$) on the total active power, obviously, does not hold anymore when we deploy Eq. (S8) to homogenize the metamaterial, since now $M$ also appears on the denominator of the expression for the effective permeability.

Before we proceed into furnishing a plausible explanation for the occurrence of regions with Im{$\mu_1$} + Im{$\mu_2$} > 0 in the case of 'tightly coupled' electric meshes, it should, at this point, be reiterated that the mechanism for overcoming losses and producing magnetic gain in our structure has a *local* origin, i.e. it occurs because we are able to design the meshes in *each* unit cell so that their current can be reversed. This reversal does *not* occur with the aid of the incident magnetic field *perpendicular* to the mesh, but with the aid of the incident magnetic field *parallel* to the mesh whose current is reversed (and perpendicular to this mesh's partner). As a result, an electric mesh whose current has been reversed can, within *each* individual unit cell, amplify a magnetic flux density component perpendicular to it. The working principle of our scheme does not rely at all on (destructive) interference amongst the cells to eliminate losses, i.e. it does not have a 'global' origin, but a much stronger 'local' one, occurring at the unit-cell level. It follows that regardless of how strongly or weakly do the unit cells interact with each other or the precise form of their interaction, the collective (effective) medium must necessarily also exhibit zero absorption of the magnetic flux density component in a specified direction (and also Im{$\mu_1$} + Im{$\mu_2$} < 0, so that the conservation of the magnetic energy is obeyed). It follows that the reason for the presence of a frequency region wherein Im{$\mu_1$} + Im{$\mu_2$} > 0 in the case of tightly coupled meshes should be traced to the limitations of Eq. (S8) in describing such a magnetic metamaterial.

Such an explanation, indeed, becomes plausible when one considers the precise methodology that should be followed in assigning bulk electromagnetic parameters in a medium. The local quasi-static electromagnetic field components will arise owing to, both, the incident field *and* the field *scattered* and/or induced by the neighbouring metaparticles. For the methodology to be self-consistent, one should start by computing the electric and magnetic multipoles induced by the incident field [S18]. In doing so, one expands the electric and magnetic dipoles in terms of the incident electromagnetic field components and their derivatives, and the electric quadrupole tensor in terms of the incident electric field. One then proceeds by determining the multipolar coefficients in the previous expansion by deploying, e.g., time-dependent quantum perturbation theory. In our case, the determination of the scattered fields should also take into account the electric connectivity of the equivalent meshes which, as we saw, results in (active and reactive) power being exchanged between them. However, the most important point for our discussion here is that, unless the unit cells are very weakly interacting, the higher-order multipolar terms cannot be ignored, even in the long-wavelength, quasi-static regime [S19]. Moreover, it turns out that these terms further depend on the origin of our coordinate system, a point that requires careful treatment before meaningful effective-medium parameters can be assigned. Such a methodology should – according to what was explained above – result in the summation $\mathrm{Im}\{\mu_1\} + \mathrm{Im}\{\mu_2\} < 0$ at every frequency point, as is the case with the 'isolated' cells. Here, however, it is the prospect of overcoming losses in metamaterials that is investigated, leaving the detailed development of a more appropriate homogenisation methodology to be the subject of a future work.

## 7. Systems with *M* degrees of freedom

The notion of so called systems having $M$ degrees of freedom is one that is very frequently encountered in diverse realms of science, from civil or mechanical engineering to quantum mechanics. In these situations, the properties (e.g., movement or oscillation) of a multi-degrees-of-freedom system in the real or in the phase space are described with the aid of $M$ independent parameters, $u_1$–$u_M$, known as *generalized Lagrange coordinates* of the system. The number of these parameters (or "coordinates") depends upon the particular form or structure of the system, on the way it is excited, as well as on the required accuracy. Generally, increasing the number of the independent parameters in the description of a system also increases the accuracy of the obtained results. As a result, there is only a limited number of cases where, e.g., an infinite-degrees-of-freedom system (also know as a *continuous* system), such as a transmission line, is described in terms of an 1-DEG equivalent system. By contrast, with the aid of a relatively small number of independent parameters, the description of a continuous system in terms of an *M*-DEG system can normally be accommodated with sufficient accuracy.

Figure S5 schematically illustrates an example of a 3-DEG system, frequently encountered in the realms of, e.g., civil or mechanical engineering. It shows the three main modes of oscillation of 3-DEG oscillator composed of three masses, $m_1 - m_3$, which are elastically attached into a vertical pole. The latter may, e.g., be simulating a concrete rod in a building or a shaft joining parts of a machine. It turns out that the 3×3 matrix describing the movement of this system [S20] is remarkably similar in its structure to the matrix resulting from the application of the 'mesh current method' to an electrical circuit of three coupled *RLC* meshes. This is simply owing to the well-known analogy between mechanical and electrical oscillators, which is also applicable in the case of infinite-degree-of-freedom (continuous) systems.

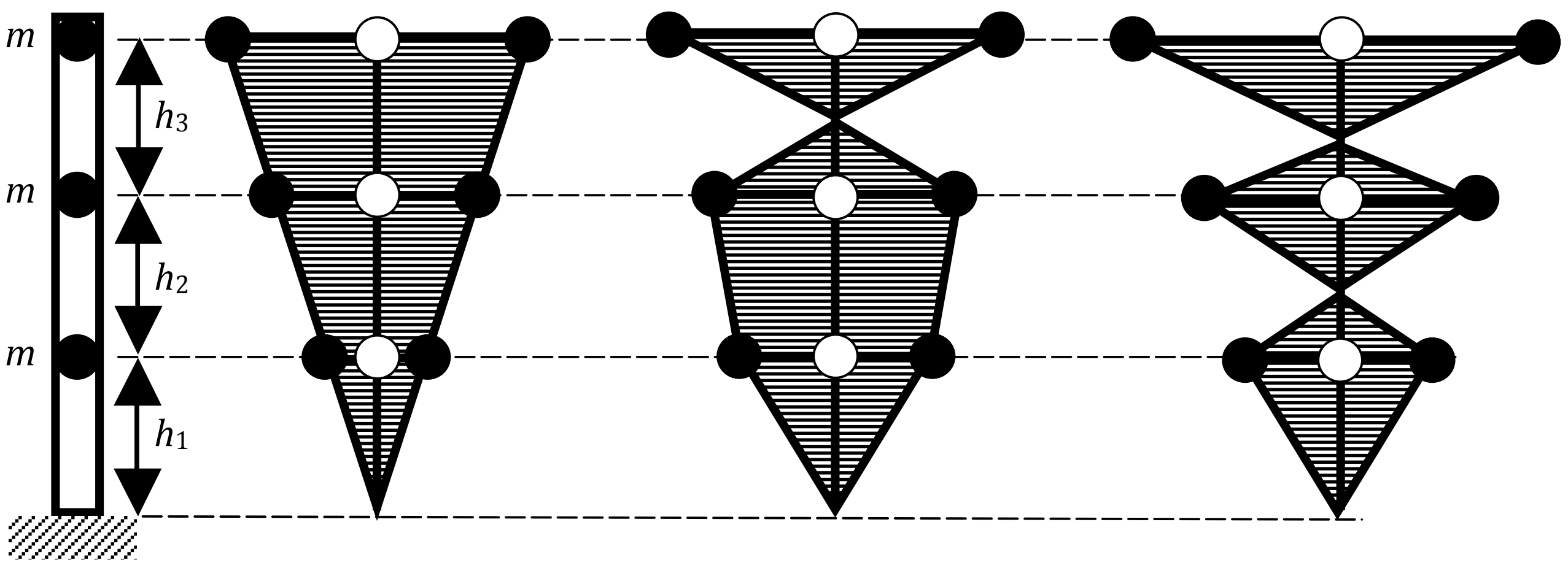

**Figure S5:** Main modes of oscillation of a 3-DEG vertical rod oscillator.

The system that we introduced in this work consists of two electrically (capacitively) connected meshes; hence, it is described in terms of two independent parameters ($I_1$ and $I_2$) and the resulting (square) matrix has order 2×2, i.e. the system is a 2-DEG one. This arrangement enabled lossless propagation for one component of the magnetic flux density (propagating along any of the allowed two directions, orthogonal to the direction of this component) at a given frequency region – actually, as we saw in Figs. 4 and S3, the propagation can, indeed, be made lossless for *two* magnetic flux density components, but this occurs at different frequency regions for each component.

It is possible, however, to envisage a system with a larger number of meshes being electrically connected, e.g., a system wherein each mesh resides at a separate plane in a cubical cell, i.e. a 6-DEG system. Depending on the design, the structure may, of course, posses even more degrees of freedom (e.g., a 12-DEG system) if two or more meshes are electrically connected in each side of the unit cell. We speculate that with a judicious choice of the electrical ($R$, $L$, $C$) parameters, such a *discrete* (i.e., non-continuous) $M$-DEG system may enable lossless propagation for two out of the three orthogonal components of the magnetic flux density, in the *same* frequency region. This can occur as long as there is sufficient imbalance in the values of the resistances of the meshes to allow for active power to flow away from the meshes residing at a plane (e.g., the *xy*-plane) towards the meshes residing at the other two orthogonal planes (e.g., the *xz*- and *yz*-planes). In this manner, the meshes at the *xy*- plane will, at the same frequency region, act as 'sources' of electrical power, i.e. they will remit active power to the meshes residing at the *xz*- and *yz*-planes – possibly to the point of *reversing* the current circulating the meshes at the *xz*- and *yz*-planes. It follows that the imaginary parts of the magnetic susceptibilities and permeabilities associated with the meshes at the *xz*- and *yz*-planes will *both* be positive in the aforementioned frequency region (or regions), i.e. the structure will be magnetically active for, *both*, the magnetic flux density components perpendicular to the *xz*- *and yz*-plane.

Let us also briefly remark that if for a certain application an *isotropic* magnetic metamaterial is required, one can use the configuration studied in Fig. 2, where both meshes reside on the same plane. To make the metamaterial isotropic one should simply place each pair of meshes on a different plane of a cube, as was shown in [25] for the case of 1-DEG split-ring resonators (SRRs). Such a configuration will exhibit the same *ultralow-loss* behaviour as the scheme of Fig. 2, while its resonant frequency will be slightly shifted compared with that of Fig. 2. It is important to realise, however, that such an *isotropic* configuration can never be characterised by $\mathrm{Im}\{\mu\} \geq 0$ (zero-loss or active) since, as was explained in the main article, the active powers associated with the individual meshes in each pair will have to be added, and the resulting *total* active power for each pair of meshes (at each plane) will always be positive, i.e. there will always be overall consumption of active power (at the resistances) for each pair of meshes. Thus, the only means of making the magnetic metamaterial lossless or active is by making it *anisotropic* – as was shown in Figs. 3 and S3.

Finally, it should be emphasised that the present scheme requires that the meshes in each unit cell are *not electrically connected* to the meshes of their neighbouring cells, i.e. the system should be *discrete* (not continuous) at the unit-cell level. If the meshes of neighbouring unit cells are electrically connected then, in the long-wavelength regime, the system becomes a continuous one (such as, e.g., a backward transmission line). These systems have been well-studied in the recent past [S21]-[S23] and, though they allow for the attainment of broadband and relatively low-loss metamaterials, they do not exhibit regions of positive imaginary parts for their effective electromagnetic parameters.

## Supplementary Notes